\documentclass[onecolumn,10pt]{aa}
\usepackage{graphicx}
\begin{document}
\title{The Everchanging Pulsating White Dwarf GD358}
\author{S.O. Kepler\inst{1}
\and R. Edward Nather\inst{2}
\and Don E. Winget\inst{2}
\and Atsuko Nitta\inst{3}
\and S. J. Kleinman\inst{3}
\and Travis Metcalfe\inst{2,4}
\and Kazuhiro Sekiguchi \inst{5}
\and Jiang Xiaojun \inst{6}
\and Denis Sullivan \inst{7}
\and Tiri Sullivan \inst{7}
\and Rimvydas Janulis \inst{8}
\and Edmund Meistas \inst{8}
\and Romualdas Kalytis \inst{8}
\and Jurek Krzesinski \inst{9}
\and Waldemar Og{\l}oza \inst{9}
\and Staszek Zola \inst{10}
\and Darragh O'Donoghue \inst{11}
\and Encarni Romero-Colmenero \inst{11}
\and Peter Martinez \inst{11}
\and Stefan Dreizler \inst{12}
\and Jochen Deetjen \inst{12}
\and Thorsten Nagel \inst{12}
\and Sonja L. Schuh \inst{12}
\and Gerard Vauclair \inst{13}
\and Fu Jian Ning \inst{13}
\and Michel Chevreton \inst{14}
\and Jan-Erik Solheim \inst{15}
\and Jose M. Gonzalez Perez \inst{15}
\and Frank Johannessen \inst{15}
\and Antonio Kanaan \inst{16}
\and Jos\'e Eduardo Costa \inst{1}
\and Alex Fabiano Murillo Costa \inst{1}
\and Matt A. Wood \inst{17}
\and Nicole Silvestri \inst{17}
\and T.J. Ahrens \inst{17}
\and Aaron Kyle Jones \inst{18,*}
\and Ansley E. Collins \inst{19,*}
\and Martha Boyer \inst{20,*}
\and J. S. Shaw \inst{21}
\and Anjum Mukadam \inst{2}
\and Eric W. Klumpe  \inst{22}
\and Jesse Larrison \inst{22}
\and Steve Kawaler \inst{23}
\and Reed Riddle \inst{23}
\and Ana Ulla \inst{24}
\and Paul Bradley \inst{25}
}
\institute{Instituto de F\'{\i}sica da UFRGS, Porto Alegre, RS - Brazil
\email{kepler@if.ufrgs.br}
\and Department of Astronomy \& McDonald Observatory, University of Texas,
Austin, TX 78712, USA
\and Sloan Digital Sky Survey, Apache Pt. Observatory, P.O. Box 59,
Sunspot, NM 88349, USA
\and Harvard-Smithsonian Center for Astrophysics,
   60 Garden Street, Cambridge, MA 02138 USA
\email {travis@whitedwarf.org}
\and Subaru National Astronomical Observatory of Japan
\email {kaz@subaru.naoj.org}
\and Beijing Astronomical Observatory,
    Academy of Sciences, Beijing 100080, China
\email {jiang@astro.as.utexas.edu}
\and University of Victoria, Wellington, New Zealand
\and Institute of Theoretical Physics and Astronomy,
    Gostauto 12,
    Vilnius 2600, Lithuania
\and Mt. Suhora Observatory,
    Cracow Pedagogical University,
    Ul. Podchorazych 2, 30-084 Cracow, Poland
\and Jagiellonian University, Krakow, Poland
\email {zola@oa.uj..edu.pl}
\and South African Astronomical Observatory
\and Universitat T\"ubingen, Germany
\and Universit\'e Paul Sabatier, Observatoire Midi-Pyr\'en\'ees,
   CNRS/UMR5572,
  14 av. E. Belin, 31400 Toulouse, France
\and Observatoire de Paris-Meudon, DAEC,
   92195 Meudon, France
\email {chevreton@obspm.fr}
\and Institutt for fysikk, 9037 Tromso, Norway
\and Departamento de F\'{\i}sica, Universidade Federal de Santa Catarina,
CP 476, CEP 88040-900, Florian\'opolis, Brazil,
\email {kanaan@fsc.ufsc.br}
\and Dept. of Physics and Space Sciences \& The SARA Observatory,
	Florida Institute of Technology, Melbourne, FL 32901
	\thanks{Southeastern
	Association for Research in Astronomy
	(SARA) NSF-REU Student.}
\and University of Florida,
	202 Nuclear Sciences Center
	Gainesville, FL  32611-8300
\and Johnson Space Center,
	2101 NASA Road 1,
	Mail Code GT2,
	Houston, TX 77058, USA
\and University of Minnesota,
	Department of Physics \& Astronomy,
	116 Church St. S.E.,
	Minneapolis, MN 55455
\and University of Georgia at Athens,
	Department of Physics and Astronomy,
	Athens, GA 30602-2451, USA
\and Middle Tennessee State University,
Department of Physics and Astronomy
Murfreesboro, TN  37132,
USA
\and Department of Physics and Astronomy,
      Iowa State University, Ames, IA 50011, USA
\and Universidade de Vigo,
Depto. de Fisica Aplicada,
Facultade de Ciencias, Campus Marcosende-Lagoas,
36200 Vigo (Pontevedra),
Spain
\email{ulla@uvigo.es}
\and Los Alamos National Laboratory,
   X-2, MS T-085
   Los Alamos, NM 87545, USA
}
\authorrunning{Kepler et al.}

\abstract{
We report 323 hours of
nearly uninterrupted time series photometric observations of the DBV
star GD~358
acquired with the Whole Earth Telescope (WET) during May 23rd to
June 8th, 2000.
We acquired more than 232\,000 independent measurements.
We also report on 48 hours of time-series photometric observations
in Aug 1996.
We detected the non-radial g-modes consistent with degree $\ell=1$ and
radial order 8 to 20 and their linear combinations up to 6th order.
We also detect, for the first time, a high amplitude
$\ell=2$ mode, with a period of 796~s.
In the 2000 WET data, the largest amplitude modes are similar to
those detected with the
WET observations of 1990 and 1994, but
the highest combination order previously detected was 4th order.
At one point during the
1996 observations, most of the pulsation energy was transferred into the
radial order $k=8$ mode, which
displayed a sinusoidal pulse shape
in spite of the large amplitude.
The multiplet structure of the individual modes changes from
year to year, and during the 2000 observations only the $k=9$ mode
displays clear normal triplet structure.
Even though the pulsation amplitudes change on timescales of days and years,
the eigenfrequencies remain essentially the same,
showing the stellar structure is not changing on
any dynamical timescale.
}
\offprints{S.O. Kepler, \email{kepler@if.ufrgs.br}}
\date{Received 6 Dec 2002 / Accepted 22 Jan 2003}
\maketitle


\section{Introduction}
GD~358, also called V777 Herculis, is the prototype of the DBV
class of white dwarf pulsators.
It was the first pulsating star detected based on a theoretical
prediction (Winget et al. 1985), and
is the pulsating star
with the largest number of periodicities detected
after the Sun.
Detecting as many modes as possible is important, as each periodicity
detected yields an independent constraint on the star's
structure.
The study of pulsating white dwarf stars has
allowed us to measure the stellar mass and composition layers,
to probe the physics at high densities, including crystallization,
and has provided a chronometer to measure the age of the oldest stars and
consequently, the age of the Galaxy.

Robinson, Kepler \& Nather (1982) and Kepler (1984) demonstrated that
the variable white dwarf stars pulsate in non-radial gravity modes.
Beauchamp et al. (1999) studied the spectra of the pulsating DBs
to determine their instability strip at
$22\,400\leq T_{\mathrm{eff}} \leq 27\,800$~K, and found
$T_{\mathrm{eff}}=24\,900$~K, $\log g=7.91$ for
the brightest DBV,
GD~358 (V=13.85),
assuming no photospheric H, as confirmed by Provencal et al. (2000).
Provencal et al. studied the HST and EUVE spectra, deriving
$T_{\mathrm{eff}}=27\,000\pm 1\,000$~K,  finding traces of
carbon in the atmosphere [$\log({\mathrm{C/He}})=-5.9\pm 0.3]$ and
a broadening corresponding to
$v\sin i=60\pm 6$~km/s. They also detected Ly$\alpha$
that is probably interstellar.
Althaus \& Benvenuto (1997) demonstrated that the
Canuto, Goldman, \& Mazzitelli (1996, hereafter CGM) self consistent theory of
turbulent convection is consistent with the $T_{\mathrm{eff}}\simeq 27\,000$~K
determination,
as GD~358 defines the blue edge of the DBV instability strip.
Shipman et al. (2002) extended the blue edge of the DBV instability strip
by finding that the even hotter star PG0112+104 is a pulsator.

Winget et al. (1994) reported
on the analysis of 154 hours of nearly continuous time series
photometry on  GD~358, obtained during
the Whole Earth Telescope (WET) run of May 1990.  The Fourier
temporal spectrum
of the light curve is dominated by periodicities in the range
1000 -- 2400~$\mu$Hz,
with more than 180 significant peaks.
They identify all of the triplet frequencies as
having degree $\ell = 1$ and, from
the details of their triplet ($k$) spacings, from which
Bradley \& Winget (1994)
derived the total stellar mass
as $0.61 \pm 0.03$ $M_\odot$, the mass of the outer helium envelope as
$2.0 \pm 1.0 \times 10^{-6}$ $M_*$, the luminosity as
$0.050 \pm 0.012~L_\odot$ and,
deriving a temperature and bolometric correction,
the distance as $42 \pm 3$~pc.
Winget et al. (1994) found changes in the $m$ spacings among the
triplet modes, and by assuming the rotational splitting coefficient
$C_{{\ell},k}(r)$ depends only on
radial overtone $k$ and the
rotation angular velocity $\Omega (r)$, interpret the
observed spacing as strong evidence of radial differential rotation,
with the outer envelope rotating some $1.8$ times faster than the core.
However, Kawaler, Sekii, {\&} Gough (1999) find that the core rotates
  faster than the envelope when they perform rotational splitting
  inversions of the observational data. The apparently contradictory
  result is due to the presence of mode trapping and the behavior of the
  rotational splitting kernel in the core of the model.
Winget et al. also found
significant power at the sums and differences of the dominant frequencies,
indicating that non--linear processes are significant, but with
a richness and complexity that rules out
resonant mode coupling as a major cause.

We show that
in the WET data set reported here
(acquired in 2000), only 12 of the periodicities can be identified as
independent g-mode pulsations,
probably all different radial overtones ($k$) with same spherical
degree $\ell=1$,
plus the azimuthal $m$ components for $k=8$ and 9. 
The high amplitude with a period of 796~s is identified as an
$\ell=2$ mode; it was not present in the previous data sets.
Most, if not all, of the
remaining periodicities are linear
combination peaks of these pulsations.
Considering there are many more observed combination frequencies
than available eigenmodes,
we interpret the linear
combination peaks as caused by non-linear effects, not real pulsations.
This interpretation is consistent with the proposal by Brickhill
(1992) and Wu (2001) that the combination frequencies appear by the non-linear
response of the medium.
Recently, van Kerkwijk et al. (2000) and Clemens et al. (2000) show that
most linear combination peaks for the DAV G29-38 do not show any
velocity variations, while the eigenmodes do.
However, Thompson et al. (2003)
argue that some combination peaks do show velocity variations.

As a clear demonstration of the power of asteroseismology,
Metcalfe, Winget, \& Charbonneau (2001)
and Metcalfe, Salaris, \& Winget (2002)
used GD~358 observed periods from Winget et~al. (1994)
and a genetic algorithm to search for
the optimum theoretical model
with static diffusion envelopes,
and constrained the $^{12}C(\alpha,\gamma)^{16}O$
cross section, a crucial parameter for many fields in astrophysics
and difficult to constrain in terrestrial laboratories.
Montgomery, Metcalfe, \& Winget (2001)
also used the observed pulsations
to constrain the diffusion of $^3{\mathrm{He}}$ in white dwarf stars.
They show their best model for GD~358 has O/C/$^4\mathrm{He}/^3\mathrm{He}$
structure, $T_{\mathrm{eff}}=22\,300 \pm 500$~K, $M_*=0.630\pm 0.015~M_\odot$,
a thick He layer,
$\log M\left({^4{\mathrm{He}}}\right)/M_* = \left(-2.79\pm 0.06\right)$,
distinct from the thin layer,
$\log M\left({^4{\mathrm{He}}}\right)/M_*  =
\left(-5.70\pm{^{+0.18}_{-0.30}}\right)$,
proposed by
Bradley \& Winget (1994).
Montgomery, Metcalfe, \& Winget's
model had
$\log M\left({^3{\mathrm{He}}}\right)/M_*  = \left(-7.49 \pm 0.12\right)$,
but Wolff et al. (2002) did not detect any $^3\mathrm{He}$
in the spectra of all the DBs they observed.
On the other hand,
Dehner \& Kawaler (1995), 
Brassard \& Fontaine (2002), and 
Fontaine \& Brassard (2002) show that a thin helium
envelope is consistent with the evolutionary models starting
at PG1159 models and ending as DQs, as diffusion is still
ongoing around 25\,000K and lower temperatures.
Therefore there could be two transition
zones in the envelope, one between the He envelope and the He/C/O layer,
where diffusion is still separating the elements, and another
transition between this layer and the C/O core.

Gautschy \& Althaus (2002) calculated nonadiabatic pulsation
properties of DB pulsators using evolutionary
models including the CGM full-spectrum
turbulence theory of convection and time-dependent element diffusion. 
They show that up to 45 dipole modes should be excited, with periods
between 335s and 2600s depending on the mass of the star, though
their models did not include pulsation--convection coupling.
They obtain a trapping-cycle length of $\Delta k=5\rightarrow 7$,
and the quadrupole modes showed instabilities comparable to the
dipole modes.

Buchler, Goupil, \& Hansen (1997) show that if there is a resonance
between pulsation modes,
even if the mode is stable, its amplitude will be necessarily nonzero.
They also point out that in case of amplitude saturation, it is the smaller
adjacent modes that show the largest amplitude variation, not the main
modes.
However, if the combination peaks are not real modes in a physical
sense, just non-linear distortion by the medium, it
is not clear that one would have resonant (mode-coupling)
between the combination peaks and real modes.

When the Whole Earth Telescope observed GD~358 in 1990, 181
periodicities were detected, but only modes from radial order $k=8$
to 18 were identified, most of them showing triplets, consistent with
the degree $\ell=1$ identification.
In fact, the observed period spacing is consistent with the
measured parallax only if the
observed pulsations have degree $\ell=1$ (Bradley \& Winget 1994).

Vuille et al. (2000) studied the 342 hours of
Whole Earth Telescope data obtained in 1994,
showing again modes with $k=8$ to 18,
and discovered up to 4th-order cross-frequencies in the power spectra.
They compared the amplitudes and phases observed with those predicted
by the pulsation--convection interaction proposed by Brickhill (1992),
and found reasonable agreement.

Note that the number of nodes in the radial direction $k$ cannot be determined
observationally and rely on a detailed comparison of the observed periods
with those predicted by pulsation models.

\section{Observations}
We report here two data sets: 48~hr of
time series photometry acquired in August 1996, with the
journal of observations presented in Table \ref{j96}.
The second data set consists of 323 hours
acquired in May-June 2000 (see Tables~\ref{jnobs} and \ref{jnobsa} 
for the observing log).
Both of these data sets were obtained simultaneously with
time resolved STIS spectroscopy with the Hubble Space Telescope,
which will be reported elsewhere.

\begin{table}
\begin{center}
\begin{tabular}{l|l|c|r|r}
\hline
Telescope & Run & Date (UT) & Start Time(UT) & Length (s)\\
\hline
\hline
Suhora 60~cm      &   suh-55  &  Aug 10 & 23:28:00 &  8890 \\
Suhora 60~cm      &   suh-56  &  Aug 12 & 20:26:40 & 14730 \\
McDonald 210~cm  &   an-0036 &  Aug 13 &  3:06:30 & 13840 \\
Suhora 60~cm      &   suh-57  &  Aug 13 & 19:12:10 & 22050 \\
McDonald 210~cm  &   an-0038 &  Aug 14 &  3:14:10 & 12920 \\
BAO 85~cm          &   bao-0026&  Aug 14 & 13:10:00 & 13610 \\
Suhora 60~cm      &   suh-58  &  Aug 14 & 23:19:10 &  5370 \\
McDonald 210~cm  &   an-0040 &  Aug 15 &  3:04:30 & 15780 \\
McDonald 96~cm  &   an-0041 &  Aug 16 &  2:54:00 & 15700 \\
BAO 85~cm          &   bao-0027&  Aug 16 & 13:01:50 &  1250 \\
BAO 85~cm          &   bao-0028&  Aug 16 & 13:42:30 &  9330 \\
McDonald 96~cm  &   an-0042 &  Aug 17 &  5:01:50 &  6420 \\
McDonald 96~cm  &   an-0043 &  Aug 18 &  4:05:30 &  3930 \\
Suhora 60-cm      &   suh-59  &  Aug 18 & 21:02:10 & 13010 \\
McDonald 96~cm  &   an-0042 &  Aug 18 &  4:05:30 &  3930 \\
McDonald 96~cm  &   an-0043 &  Aug 19 &  2:44:40 &  2100 \\
McDonald 96~cm  &   an-0044 &  Aug 19 &  2:44:40 &  2100 \\
McDonald 96~cm  &   an-0044 &  Aug 19 &  3:47:00 &  4250 \\
McDonald 96~cm  &   an-0045 &  Aug 19 &  3:47:00 &  4250 \\
McDonald 96~cm  &   an-0046 &  Aug 19 &  4:58:30 &  7160 \\
Suhora 60~cm      &   suh-60  &  Aug 19 & 20:28:00 & 15290 \\
Suhora 60~cm      &   suh-61  &  Aug 20 & 21:04:00 & 10670 \\
\hline
\hline
\end{tabular}
\end{center}
\caption{Journal of ground-based observation for GD358 in August, 1996.
\label{j96}}
\end{table}

In 1996, the observations were obtained with
three channel time series photometry
using bi-alkali photocathodes
(Kleinman, Nather, \&  Phillips, 1996),
in Texas, China, and Poland.
During 23 May to 23 June, 2000, we
observed GD~358 mainly with two and
three channel time series photometers
using bi-alkali photocathodes
and a time resolution of 5~s.
The May-June 2002 run used 13 telescopes composing the
Whole Earth Telescope. The telescopes, ranging from
60~cm to 256~cm in diameter, were located in
Texas, Arizona, Hawaii, New Zealand, China,
Lithuania, Poland, South Africa, France, Spain, Canary Islands, and Brazil.
As the pulsations in white dwarf stars are
in phase at different wavelengths (Robinson, Kepler, \& Nather
1982), we used no filters, to maximize the detected signal.

Each run was reduced
and analyzed as
described by Nather et al. (1990) and Kepler (1993),
correcting for extinction through an estimated
local coefficient, and sky variations measured continuously on
three channel and CCD observations, or sampled frequently on two channel
photometers. The second channel of the photometer monitored a
nearby star to assure photometric conditions or
correct for small non-photometric conditions.
The CCD measurements were obtained with different cameras which are
not described in detail here. At least two comparison stars were in each frame
and allowed for differential weighted aperture photometry. The 
consecutive data
points were 10 to 30~s apart, depending if the CCDs were frame 
transfer or not.

\begin{table}
\begin{center}
\caption{Journal of ground-based observation for GD358 in May-June, 2000.\label{jnobs}}
\begin{tabular}{lllrrr}
  Run Name&
  Telescope&
  Date (UT)&
  Start Time &
  Length &\\
&&(UT)\hfil&(s)\hfil&\\
jr0523  &  Moletai 1.65m &  May 23 00 &  22:15:49 & 6360 &  \\
tsm-0074  &  McDonald 2.1m &  May 24 &  4:59:00 & 20050 &  \\
jr0524  &  Moletai 1.65m &  May 24 &  20:24:35 & 10255 &  \\
tsm-0075  &  McDonald 2.1m &  May 25 &  3:25:00 & 25440 &  \\
suh-089  &  SUHORA 0.6m &  May 25 &  23:58:30 & 4715 &  \\
tsm-0076  &  McDonald 2.1m &  May 26 &  3:18:30 & 25350 &  \\
suh-090  &  Suhora 0.6m &  May 26 &  20:07:20 & 18995 &  \\
jr0526  &  Moletai 1.65m &  May 26 &  20:26:00 & 12545 &  \\
teide02  &  Tenerife IAC 0.8m &  May 27 &  01:25:20 & 13270 &  \\
tsm-0077  &  McDonald 2.1m &  May 27 &  03:00:00 & 26100 &  \\
suh-091  &  Suhora 0.6m &  May 27 &  19:56:20 & 15240 &  \\
sa-et1  &  SAAO 0.75m &  May 28&  00:18:30 & 15220 &  \\
teide04  &  Tenerife IAC 0.8m &  May 28 &  0:37:10 & 14870 &  \\
tsm-0078  &  McDonald 2.1m &  May 28 &  2:51:00 & 29040 &  \\
teide05  &  Tenerife IAC 0.8m &  May 28 &  21:53:20 & 24700 &  \\
jr0528  &  Moletai 1.65m &  May 28 &  20:28:20 & 9550 &  \\
tsm-0079  &  McDonald 2.1m &  May 29 &  2:50:00 & 27300 &  \\
calto2905  &  Calar Alto 1.23m &  May 29 &  21:20:00 & 6000 &  CCD  \\
teide06  &  Tenerife IAC 0.8m &  May 29 &  22:01:50 & 4375 &  \\
gv-2905  &  OHP &  May 29 &  22:06:00 & 15530 &  \\
teide07  &  Tenerife IAC 0.8m &  May 29 &  23:28:00 & 11460 &  \\
teide08  &  Tenerife IAC 0.8m &  May 30 &  2:50:50 & 6850 &  \\
tsm-0080  &  McDonald 2.1m &  May 30 &  3:00:00 & 28200 &  \\
suh-092  &  Suhora 0.6m &  May 30 &  20:39:30 & 14880 &  \\
gv-2906  &  OHP 1.93m &  May 30 &  20:44:00 & 4340 &  \\
teide09  &  Tenerife IAC 0.8m &  May 30 &  21:58:50 & 25270 &  \\
sjk-0401  &  Hawaii UH 0.6m &  May 31 &  7:19:00 & 26615 &  \\
gv-2907  &  OHP, 1.93 m &  May 31 &  20:44:00 & 20400 &  \\
teide10  &  Tenerife IAC 0.8m &  May 31 &  22:25:20 & 8195 &  \\
calto3105  &  Calar Alto 1.23m &  May 31  &  21:30:00 & 5560 &  CCD  \\
sa-od033  &  SAAO 0.75m &  May 31 &  22:31:00 & 4000 &  \\
sjk-0402  &  Hawaii UH  0.6m &  Jun 1 &  6:05:00 & 31265 &  \\
jxj-0103  &  BAO 0.85m &  Jun 1 &  13:18:20 & 6950 &  \\
suh-093  &  Suhora 0.6m &  Jun 1 &  19:55:30 & 19310 &  \\
jr0601  &  Moletai 1.65m &  Jun 1 &  20:12:00 & 13145 &  \\
sa-od035  &  SAAO 0.75m &  Jun 1 &  20:22:00 & 10600 &  \\
calto0106  &  Calar Alto 1.23m &  Jun 1 &  20:46:00 & 4845 &  CCD  \\
gv-2908  &  OHP 1.93m &  Jun 1 &  20:48:00 & 19650 &  \\
teide11  &  Tenerife IAC 0.8m &  Jun 1 &  22:38:40 & 2230 &  \\
teide12  &  Tenerife IAC 0.8m &  Jun 1 &  23:16:20 & 12460 &  \\
tsm-0081  &  McDonald 2.1m &  Jun 2 &  2:50:00 & 24000 &  \\
teide13  &  Tenerife IAC 0.8m &  Jun 2 &  3:00:20 & 7180 &  \\
sara030&Sara 0.9m&Jun 2 &8:30:00&   14400 &CCD\\
sjk-0403  &  Hawaii UH 0.6m &  Jun 2 &  6:04:00 & 32505 &  \\
jxj-0003  &  BAAO 0.85m &  Jun 2 &  15:24:50 & 11030 &  \\
suh-094  &  Suhora 0.6m &  Jun 2 &  19:53:00 & 19530 &  \\
gv-2909  &  OHP 1.93m &  Jun 2 &  21:40:00 & 16750 &  \\
jr0602 &  Moletai 1.65m &  Jun 2 &  22:03:45 & 4740 &  \\
calto0602  &  Calar Alto 1.23m &  Jun 2 &  23:20:00 & 4030 &  CCD  \\
teide15 &  Tenerife IAC 0.8m &  Jun 3 &  00:22:40 & 16630 &  \\
\end{tabular}
\end{center}
\end{table}

\begin{table}
\begin{center}
\caption{Journal of ground-based observation for GD358 in May-June, 2000.
(cont.)\label{jnobsa}}
\begin{tabular}{lllrrr}
  Run Name&
  Telescope&
  Date (UT)&
  Start Time &
  Length &\\
&&(UT)\hfil&(s)\hfil&\\
calto0602.2  &  Calar Alto 1.23m &  Jun 3 &  00:35:00 & 2800 &  \\
tsm-0082 &  McDonald 2.1m &  Jun 3 &  3:03:30 & 22110 &  \\
sara031&Sara 0.9m&Jun 3& 4:15:00&   26100:30&CCD\\
sjk-0404  &  Hawaii UH 0.6m &  Jun 3 &  5:59:30 & 31270 &  \\
jxj-0105  &  BAO 0.85m &  Jun 3 &  13:27:50 & 4185 &  \\
sjk-0405  &  Hawaii UH 0.6m &  Jun 3 &  14:42:30 & 1465 &  \\
jxj-0106  &  BAO 0.85m &  Jun 3 &  16:00:20 & 8725 &  \\
suh-095  &  Suhora 0.6m &  Jun 3  &  20:02:50 & 19055 &  \\
jr0603  &  Moletai 1.65m &  Jun 3 &  20:24:55 & 12305 &  \\
sa-od037  &  SAAO 0.75m &  Jun 3 &  20:49:00 & 10655 &  \\
teide17  &  Tenerife IAC 0.8m &  Jun 4 &  00:26:00 & 16760 &  \\
sara032&Sara 0.9m&Jun 4 &4:08:00&   26940&CCD\\
sjk-0406  &  Hawaii UH 0.6m &  Jun 4 &  5:37:00 & 34055 &  \\
tsm-0083  &  McDonald 2.1m &  Jun 4 &  7:46:30 & 11310 &  \\
jxj-0107  &  BAO 0.85m &  Jun 4 &  12:32:50 & 25280 &  \\
suh-096  &  Suhora 0.6m &  Jun 4 &  20:21:00 & 16775 &  \\
sa-od039  &  SAAO 0.75m &  Jun 4 &  21:42:00 & 7110 &  \\
calto0604 &  Calar Alto 1.23m &  Jun 4 &  23:06:37 & 4290 &  CCD\\
teide19  &  Tenerife IAC 0.8m &  Jun 5 &  0:14:30 & 17270 &  \\
tsm-0084  &  McDonald 2.1m &  Jun 5 &  3:00:00 & 6150 &  \\
sara034&Sara 0.9m&Jun 5 &4:47:00&   24720&CCD\\
jxj-0108 &  BAO 0.85m &  Jun 5 &  12:33:20 & 25295 &  \\
suh-097  &  Suhora 0.6m &  Jun 5 &  20:04:00 & 18545 &  \\
jr0605{\_}1  &  Moletai 1.65m &  Jun 5 &  20:58:25 & 5555 &  \\
sa-od042  &  SAAO 0.75m &  Jun 5 &  21:49:00 & 8005 &  \\
jr0605{\_}2  &  Moletai 1.65m &  Jun 5 &  22:52:55 & 3925 &  \\
teide20  &  Tenerife IAC 0.8m &  Jun 6 &  1:08:00 & 13685 &  \\
tsm-0085  &  McDonald 2.1m &  Jun 6 &  2:55:00 & 28800 &  \\
sara035&Sara 0.9m&Jun 6 &4:08:00&   10080&CCD\\
edjoh01  &  NOT &  Jun 6 &  21:33:40 & 10150 &  \\
edjoh02  &  NOT &  Jun 7 &  1:33:10 & 13025 &  \\
teide22  &  Tenerife IAC 0.8m &  Jun 7 &  0:16:40 & 16850 &  \\
teide23  &  Tenerife IAC 0.8m &  Jun 7 &  21:00:00 & 28800 &  \\
edjoh03  &  NOT 2.5m &  Jun 7 &  22:23:40 & 15755 &  \\
suh-098 &  Suhora 0.6m &  Jun 8 &  20:05:10 & 17065 &  \\
sara036&Sara 0.9m&Jun 10 &4:41:00&  7800&CCD\\
sara037&Sara 0.9m&Jun 11 &4:01:00&   9780&CCD\\
sara038&Sara 0.9m&Jun 12 &3:57:00&   10600 &CCD\\
sara039&Sara 0.9m&Jun 20 &7:09:00&  14700&CCD\\
sara040&Sara 0.9m&Jun 21 &3:36:00&   28920&CCD\\
sara041&Sara 0.9m&Jun 22 &3:29:00&    22020&CCD\\
sara042&Sara 0.9m&Jun 23 &3:23:00&   28680&CCD\\
\end{tabular}
\end{center}
\end{table}

\begin{figure}
\centering
\includegraphics[width=\linewidth]{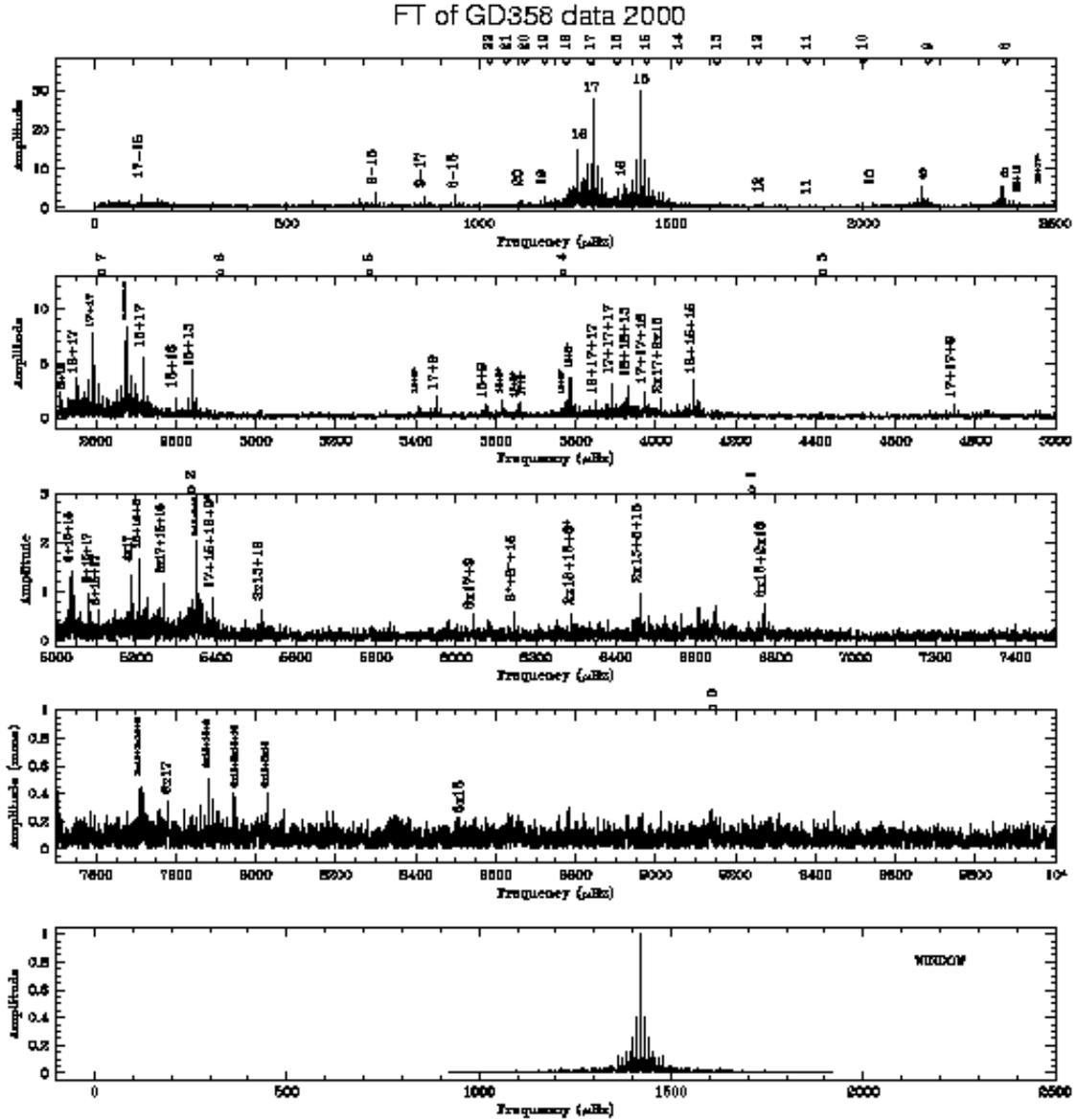}
\caption{Fourier transform of the 2000 data set. The main power is
concentrated
in the region between 1000~$\mu$Hz and 2500~$\mu$Hz. The marks on top of the
graph are the asymptotic equally spaced periods prediction, and the numbers
represent the radial order $k$ value, with Winget et al. (1994)
identification.
The vertical scale on each panel are adjusted to accommodate the large range
of amplitudes shown, and the noise level which decreases from 0.29~mma
up to 3000~$\mu$Hz to 0.19~mma upwards.
\label{dft2000}}
\end{figure}

After this preliminary reduction, we brought the
data to the same fractional amplitude
scale and converted the middle of integration times to Barycentric
Coordinated Time TCB (Standish 1998).
We then computed a Discrete Fourier Transform (DFT)
for the combined 2000
data, shown in Fig. \ref{dft2000}.
Due to poor weather conditions during the run, our coverage is not continuous,
causing gaps in the observed light curve; these gaps produce
aliases in the Fourier transform. At the bottom of Fig.
(\ref{dft2000}) we present the spectral window, the Fourier
transform of a single sinusoid sampled exactly as the real
data. It shows the pattern of peaks each individual
frequency in the data introduces in the DFT.

The Fourier spectra displayed in Fig.
(\ref{dft2000}) looks similar to the ones obtained in 1990 and 1994
(see Figure \ref{yearlydft}),
but the amplitude of all the modes changed significantly.
As we describe in more detail later, the most striking feature of the
2000 data is the absence of triplets, except for $k=9$.
The 1996 data are even more unusual than the WET runs due to the
observation of amplitude changes over an unprecedented short time.
We describe these observations in more detail in section 3.
We then describe the 2000 observations and our interpretations of them
in Sections 4 through 7.

\section{Changes in the Dominant Mode in 1996}

We observed GD~358 in August 1996 to provide simultaneous observations
to compare to HST time resolved  spectroscopy made on August 16.
The 1996 data covers 10 days of the most remarkable amplitude behavior
ever seen in a pulsating white dwarf. Observing this behavior is
serendipitous, as individual observers and the WET have observed GD~358
off and on for 20 years without seeing this sort of behavior.

Figure~\ref{gd358amp} shows how the amplitude of the $k=8$ $P=423$~s mode
changed with time during our observations in August~1996.
The amplitude changes we saw in our optical data are
unprecedented in the observations of pulsating white dwarf stars; no
report has
been made of such a large amplitude variation in such a short amount of
time.  Here we describe what we found in our data.

The lightcurves acquired in August 1996 are displayed in Fig.~\ref{gd358lc1}
and the Fourier transform for each lightcurve is shown in Figure 4.
Those in the first and second panels of
Fig.~\ref{gd358lc1} look very different from each other. The
Fourier transform of the lightcurve from the first panel is similar to
that from the 1994 WET data where we identified over 100 individual
periodicities, while the Fourier transform of the second panel is
dominated by only a single periodicity (Fig.~\ref{dftchange});
this represents a complete change in the mode structure, as well as the
period of the dominant mode, in about one day!

In run an--0034, the $k=8$ P=423~s mode's amplitude is 170~mma,
which is the largest amplitude we have ever seen for this mode.
To check for additional pulsation power (perhaps lower
amplitude pulsations dwarfed by the 423~s mode power), we prewhitened
the an--0034 lightcurve by the 423~s mode.  Prewhitening subtracts
a sinusoid with a specified amplitude, phase and period from the original
lightcurve, and it helps us look for smaller amplitude
pulsations by eliminating the alias pattern of the dominant pulsation mode
from the Fourier transform.
In Fig.~\ref{an34dft}, we show the Fourier transform of the
an--0034 lightcurve both before and after prewhitening.  We see
now that  GD~358 was indeed dominated entirely by a single mode at a
different period from the dominant mode a day earlier.
We refer to this event by the musical term ``{\it forte}'',
or more informally as the ``whoopsie''.

\begin{figure}
\includegraphics[width=\linewidth]{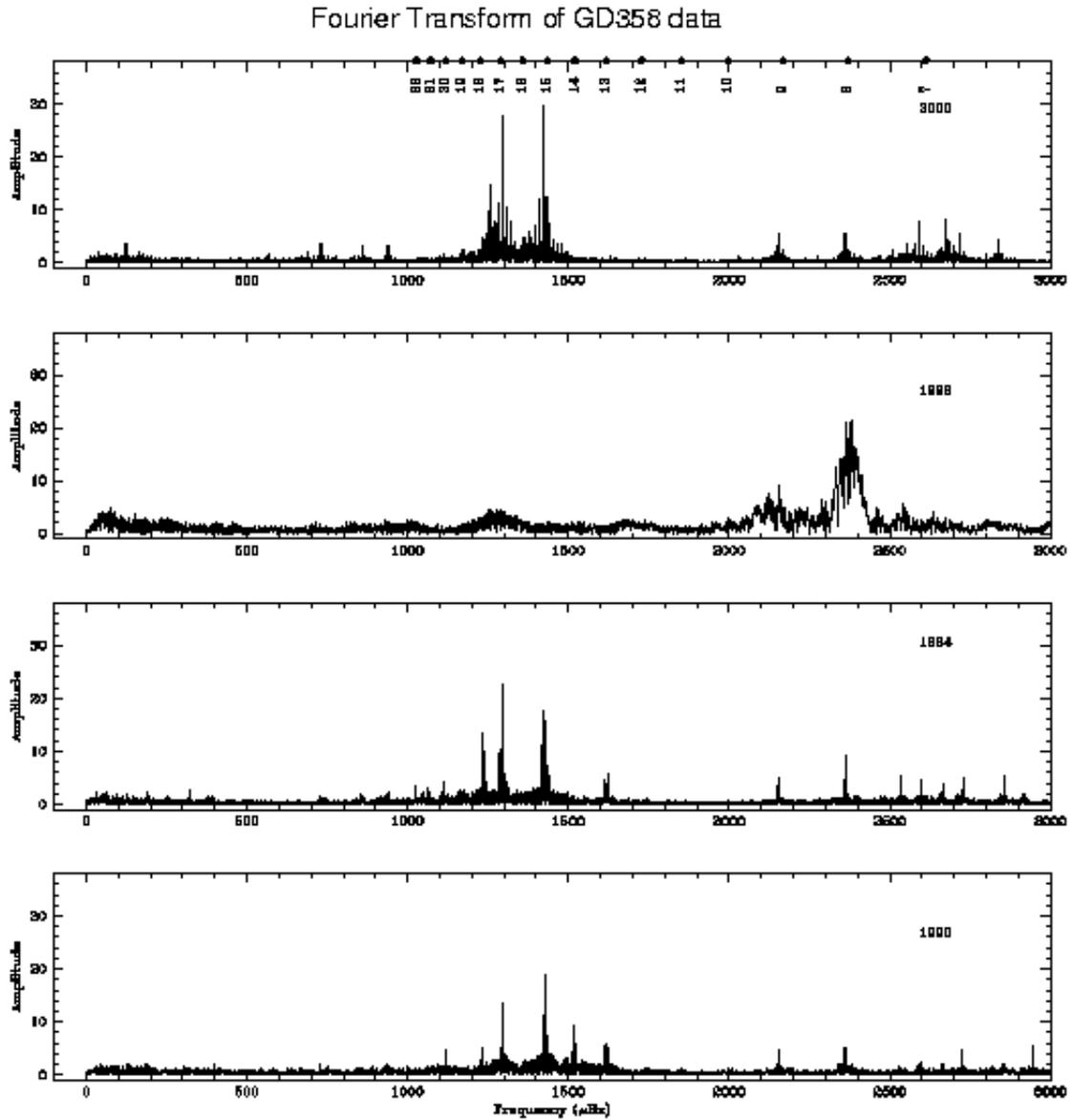}
\caption{Fourier transform of the GD~358 data, year by year.}
\label{yearlydft}
\end{figure}

\begin{figure}
\centering
\includegraphics[height=20cm]{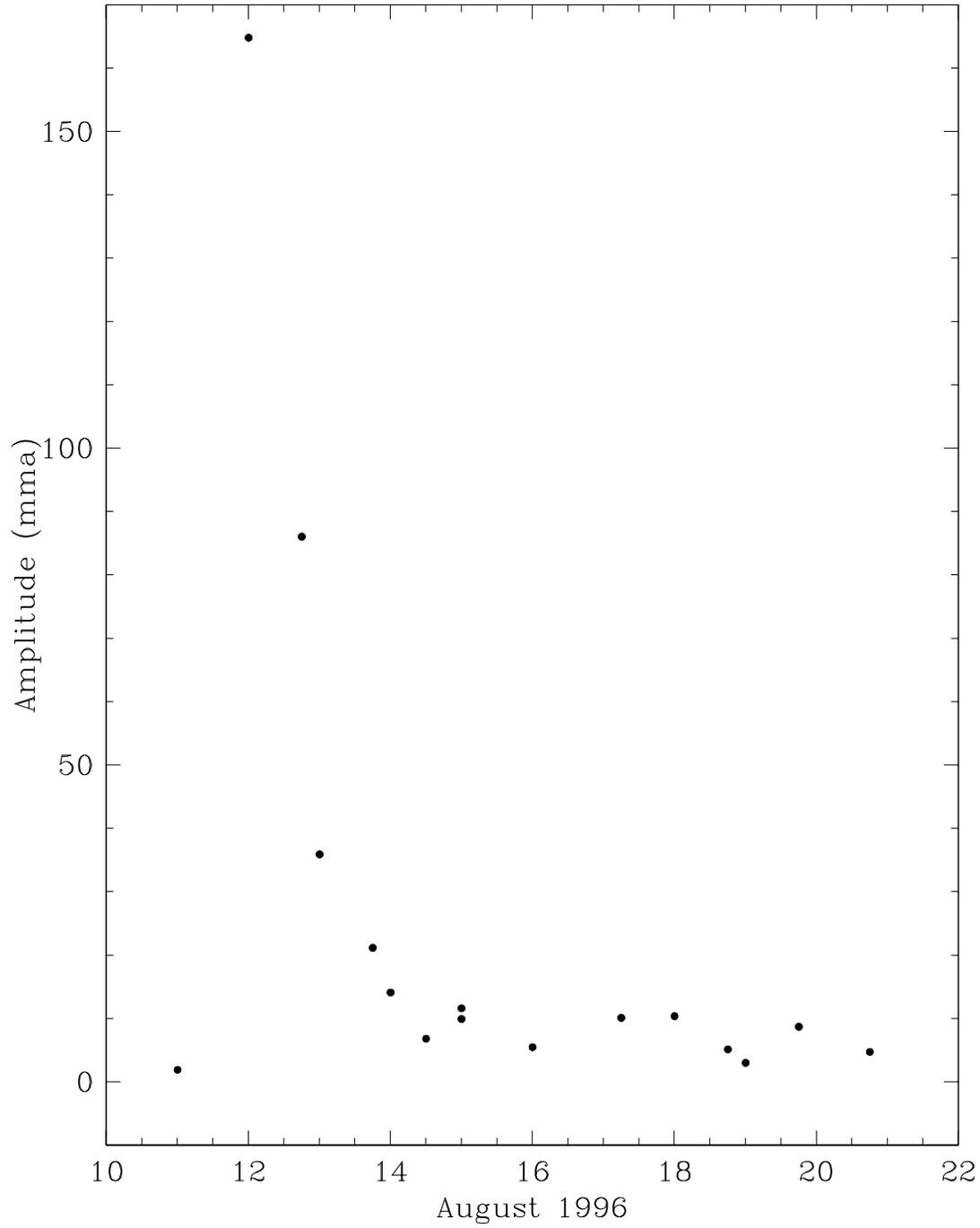}
\caption[Amplitude Modulation of 423~s Mode]
{The amplitude modulation of GD~358 423~s mode observed in the optical in
August, 1996. The timescale of this change is surprisingly short. We have
never observed such a fast amplitude modulation in any of the pulsating
white dwarf stars.}
\label{gd358amp}
\end{figure}

\begin{figure}
\centering
\includegraphics[height=20cm]{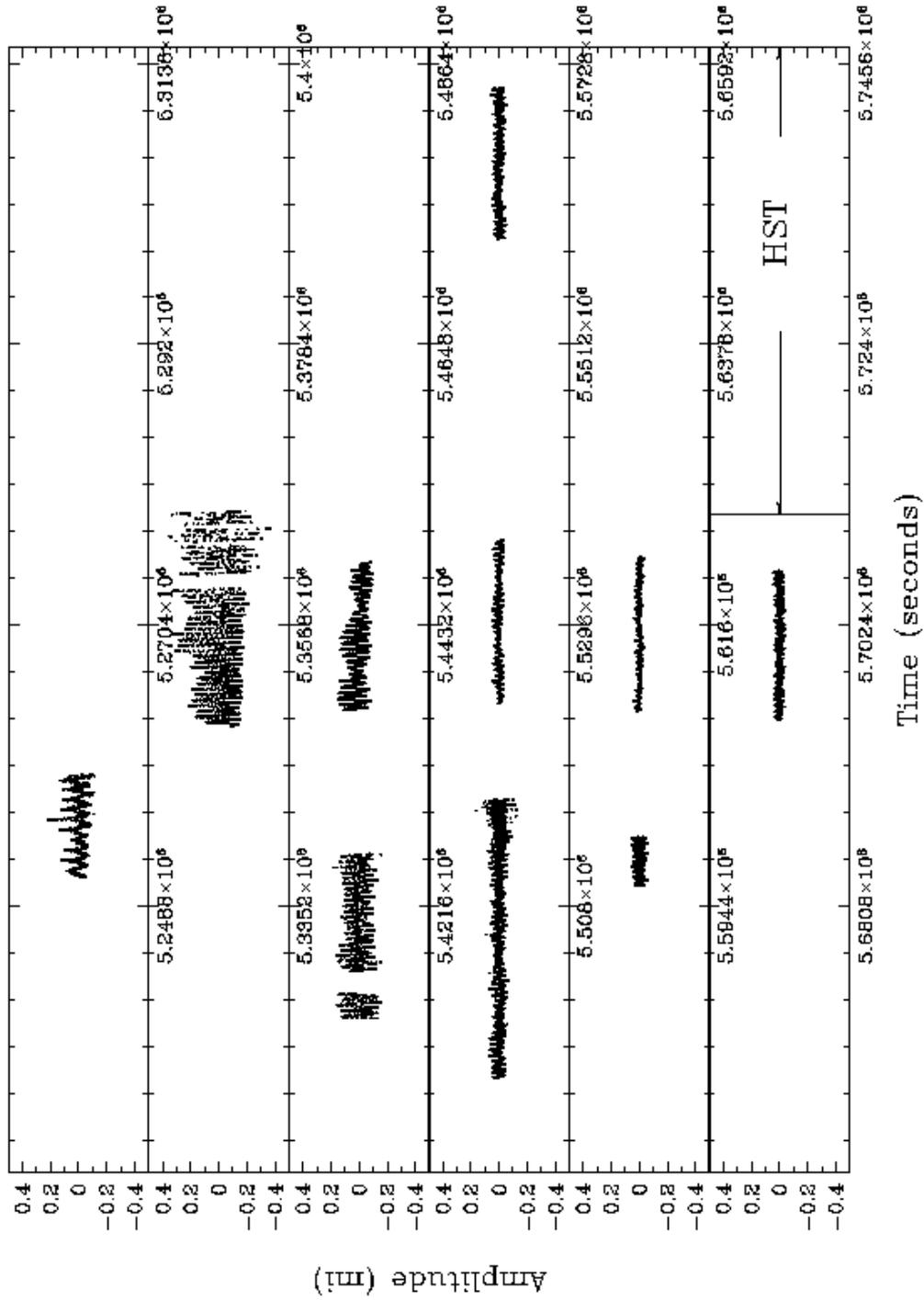}
\caption[Optical lightcurve of GD~358 in August 1996]
{First half of the ground-based optical lightcurve of GD~358 in August 1996.
Each panel is one day long.}
\label{gd358lc1}
\end{figure}

\begin{figure}
\centering
\includegraphics[height=20cm]{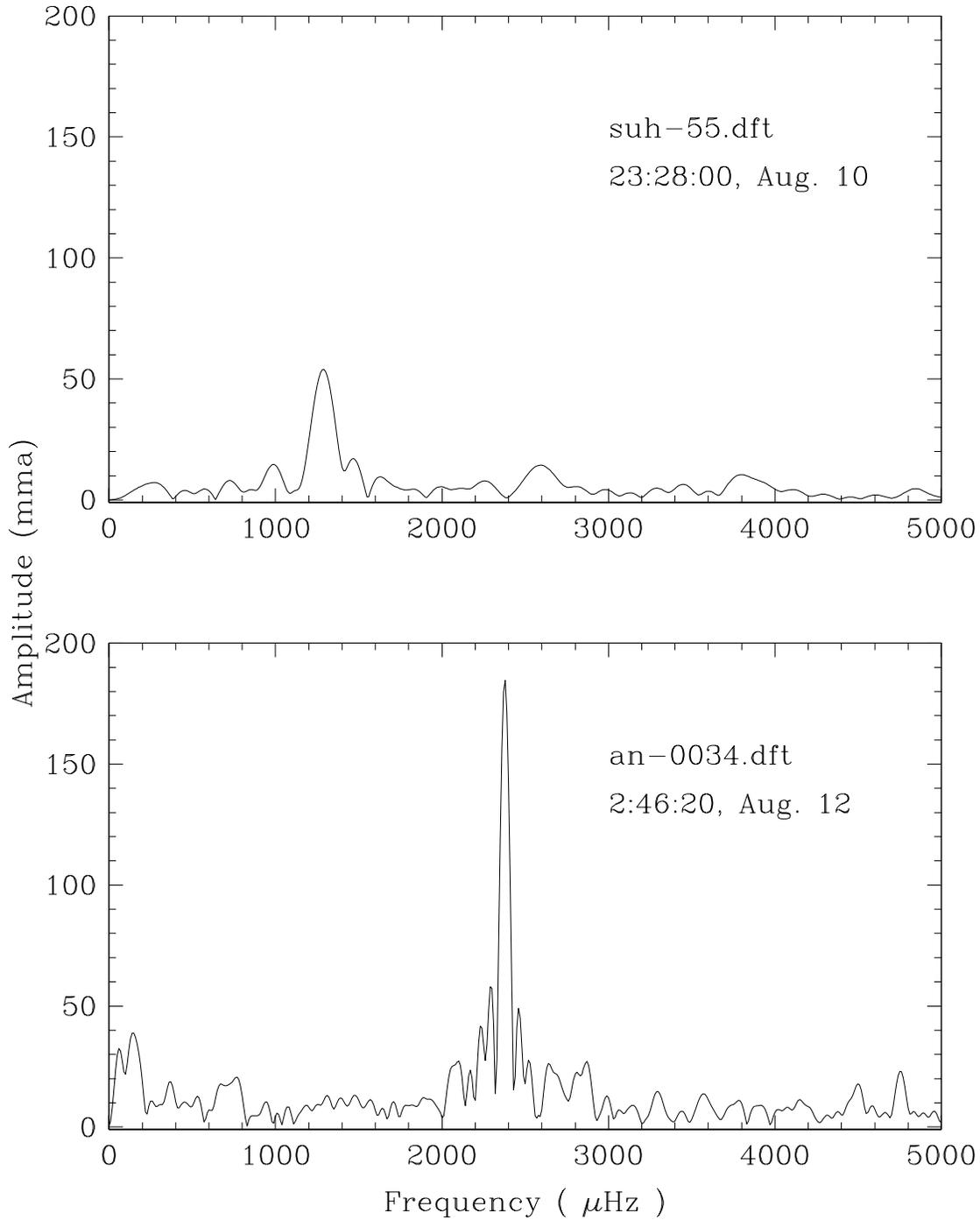}
\caption[Fourier Transforms of GD~358 on the 1st and 2nd day of the
Optical Observation]
{Fourier transforms of GD~358 observed one day apart.  The top panel shows
the Fourier transform of the data taken on the 1st day of the 3-site
campaign (suh--55: taken in Poland with start time at 23:28:00 UT on August
10), and the bottom panel shows the data taken about one day later from
McDonald (an--0034: taken with start time at 2:48:20 UT on August 12). The
observed power has shifted completely and dramatically, both in frequency
and amplitude.}
\label{dftchange}
\end{figure}

\begin{figure}
\centering
\includegraphics[totalheight=15cm,angle=270]{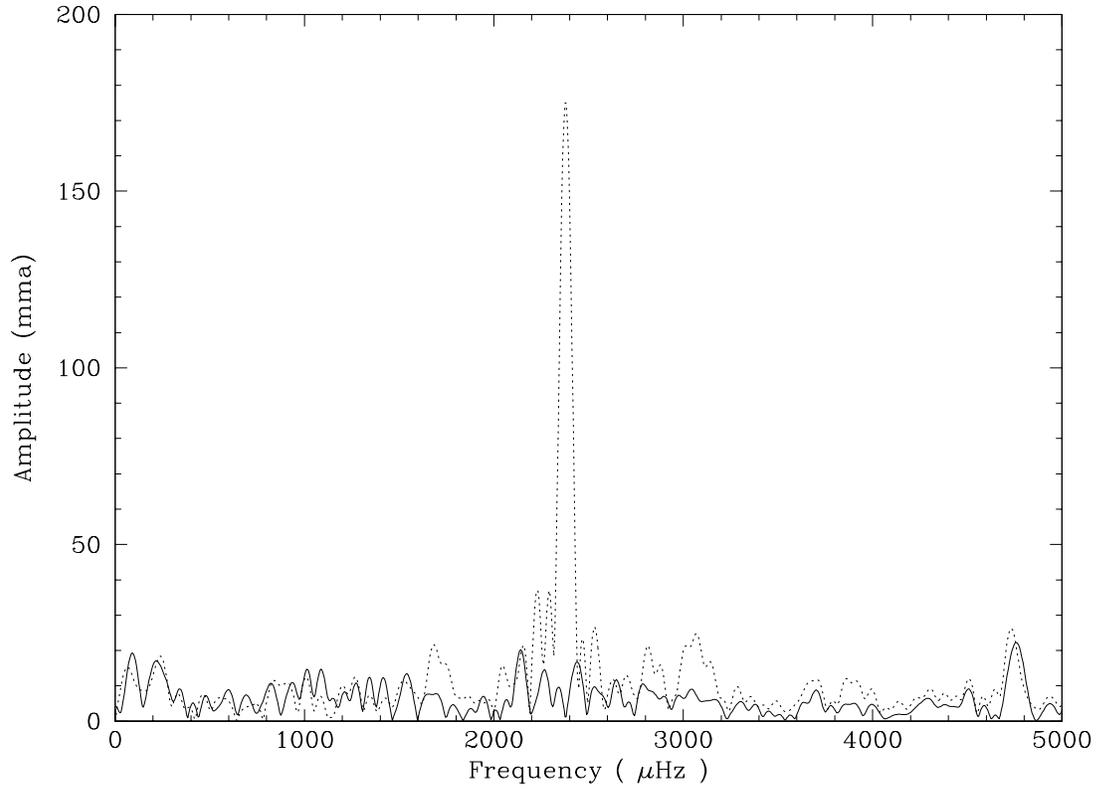}
\caption[Fourier Transform of an-0034 Before and After Prewhitening]
{ Fourier transform of an-0034 before (dotted line) and after (solid line)
it was prewhitened by the 423~s mode. After prewhitening, there is little
significant power left. The lightcurve was dominated by one mode,
a possible explanation for
why the lightcurve looked so linear (sinusoidal) in
Fig.~\ref{gd358lc1}.}
\label{an34dft}
\end{figure}

\begin{figure}
\centering
\includegraphics[height=20cm]{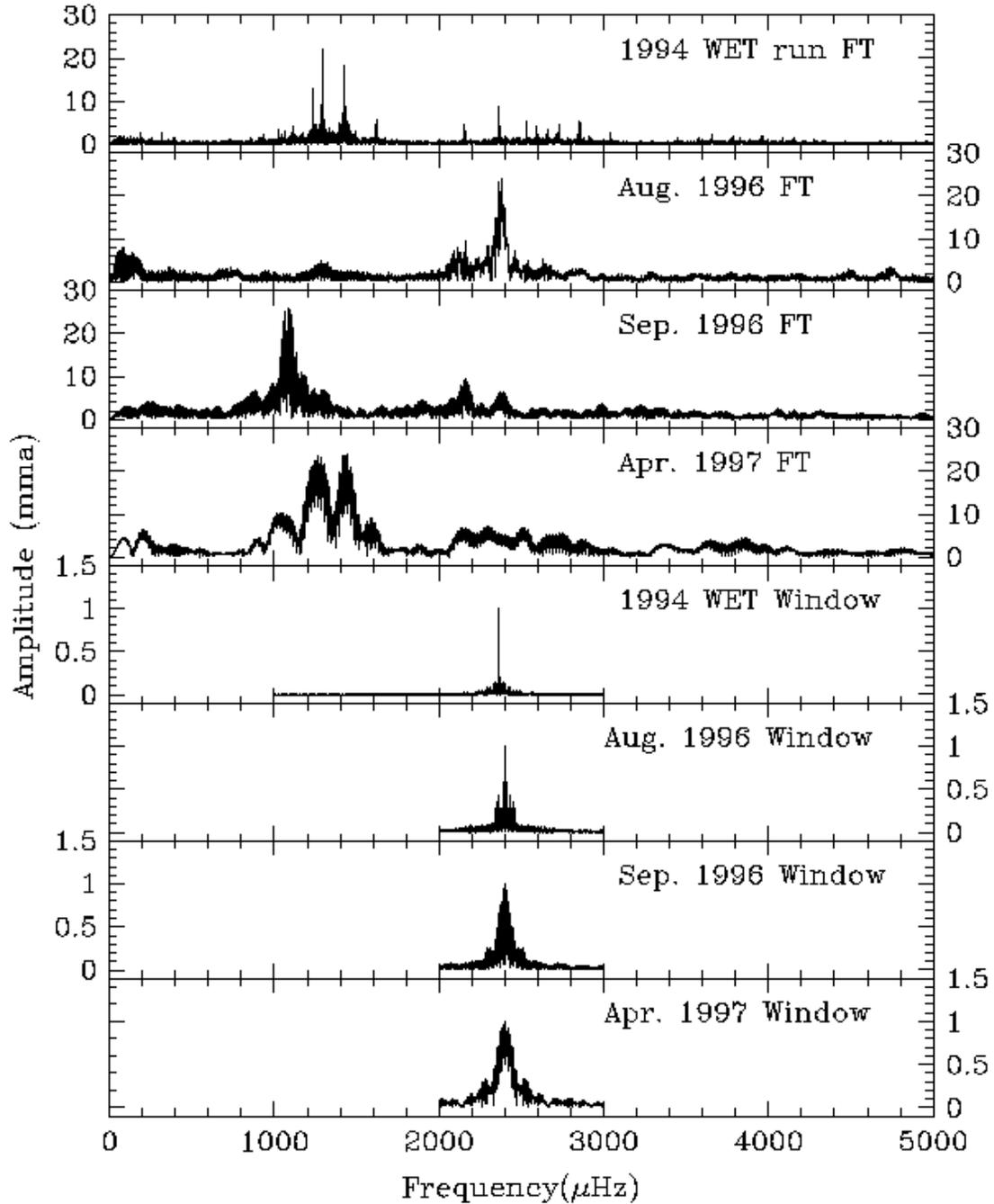}
\caption[GD~358 Fourier Transforms at 3 Different Times]
{GD~358 Fourier transform at four different times along with their
spectral windows.  The 1994 and 1997 Fourier transforms look similar
(within the observed frequency resolution, that is). The September
1996 data look similar as well to these two data sets, but the highest
amplitude modes have shorter frequencies (longer period).  Obviously,
the August 1996 Fourier transform  looks very different from the other
Fourier transforms.}
\label{fig4opt}
\end{figure}

Given the spectacular behavior of GD~358 in August 1996, we obtained
follow-up observations in September 1996 and April 1997.
Table~\ref{journal2} shows the journal of observations for the
September 1996  and April 1997 data.
The lightcurve and the power spectrum during these observations
(Fig.~\ref{fig4opt}) seem to have returned to the more or less
normal state seen in the past (Fig.~\ref{yearlydft}), not  the unusually high
amplitude state it was in in August 1996 (Fig.~\ref{dftchange}).

\begin{figure}
\centering
\includegraphics[height=20cm]{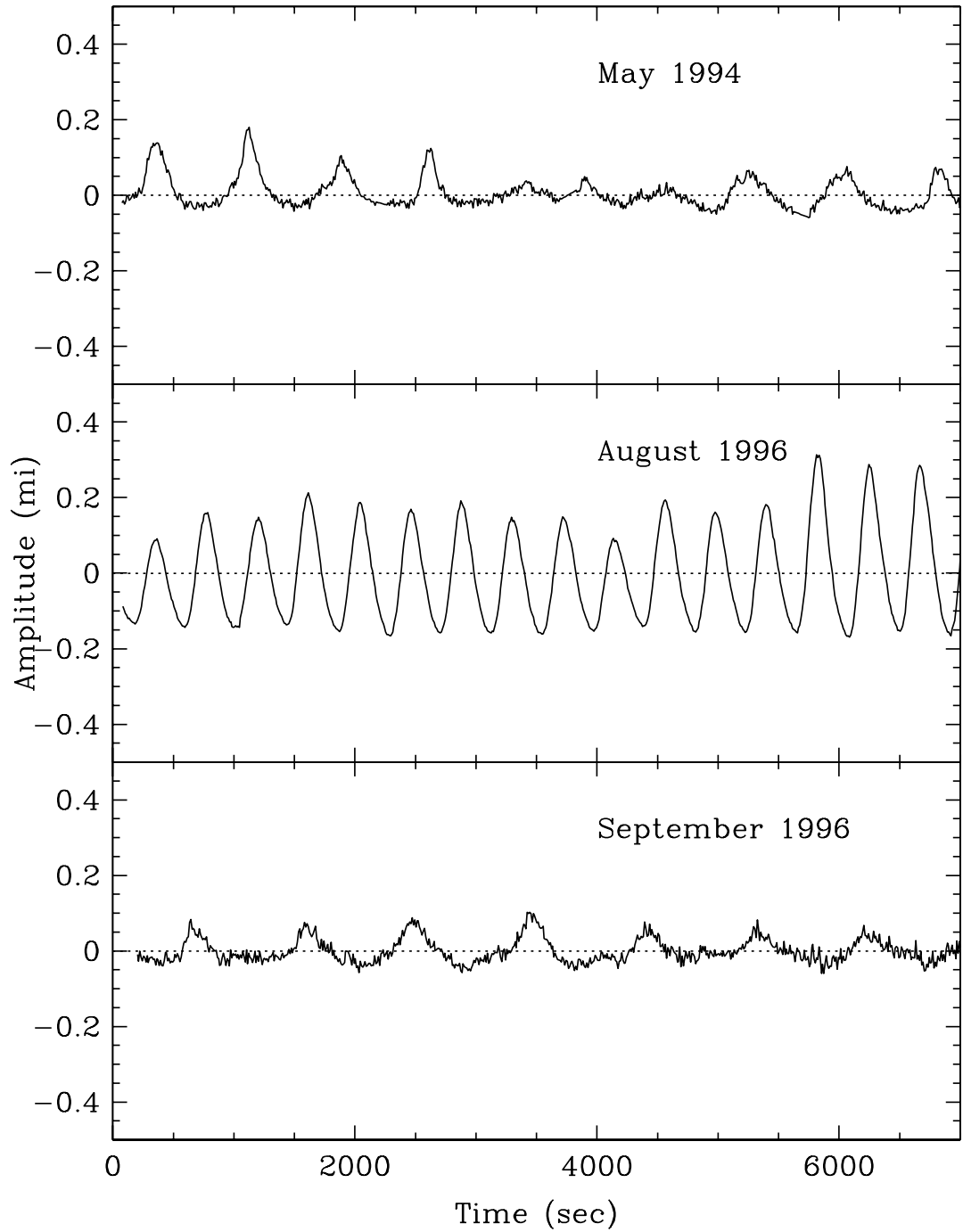}
\caption[GD~358 Lightcurves over Time]
{GD~358 lightcurves over time. The shape of the lightcurve was sinusoidal
when the amplitude was highest. The 1994 and September 1996 data exhibit
similar pulse shapes and their corresponding power spectra also look
similar (Fig.~\ref{fig4opt}).
}
\label{pulse}
\end{figure}

We show the lightcurves of GD~358 at three different times in
Fig.~\ref{pulse}.  The middle panel shows the lightcurve when the
amplitude of the 423~s mode was at its largest.  The lightcurve looks
almost sinusoidal, with the single 423~s mode in the power spectrum. 
The result of this is that we obtain similar values for the
peak-to-peak semi-amplitudes and the FT amplitude of the 423~s mode;
this implies that a single spherical harmonic is a good representation
of the stellar pulsation at this time.
The other two lightcurves,
however, each containing several pulsation modes, are less sinusoidal. If
the non-sinusoidal nature of a lightcurve comes from the fact that many
modes are present simultaneously, then one would expect the shape of
the lightcurve to be sinusoidal only when it is pulsating in a single
mode.  On the other hand, in the August 1996 sinusoidal lightcurve,
the peak--to--peak light variation was about 44\% of the star's
average light in the optical.  We would expect such a large
light variation
to introduce nonlinear effects into the lightcurve, even if the star
is pulsating in a single mode, causing the lightcurve to look
nonsinusoidal.  Thus, the nearly sinusoidal shape of our lightcurves
(Fig.~\ref{pulse}) is a mystery, except for the
theoretical models of Ising \& Koester 2001, which predict sinusoidal
shapes for large amplitude modes even with the  nonlinear response of the
envelope.

After the $P=423$~s mode reached its highest amplitude in run an-0034,
the $k=9$ $P=464$~s mode started to grow and the 423~s became smaller,
but there was still very little sign of the usually dominant $k=17$
$P=770$~s
mode.
In Fig. \ref{yearlydft}, we present the Fourier amplitude spectra of
the light curves obtained each year, 2000 on top, 1996, 1994, and 1900
on the bottom, on the same vertical scale. Note that the 1996 data set is low
resolution, because of its smaller amount of data.
It is clear that the periodicities change amplitude from
one data set to the other.
It is important to notice that the periodicities,
when present, have similar frequencies over the years. The amplitudes
change, and even subcomponents (different $m$ values) may appear and
disappear, but when they are present, they have basically the same
frequencies (typically to within 1~$\mu$Hz).

In the
September 1996 data, the Fourier transform shows that GD~358 is pulsating
in periods similar to what we are familiar with from the WET data of 1990,
although
the highest peaks are at $1082~\mu$Hz, $2175~\mu$Hz and $2391~\mu$Hz.
The very limited data set and the complex pulsating structure of the
star makes interpretation of these peaks difficult
(Fig.~\ref{fig4opt}).   It is
not until the data taken in April 1997 when we observe the 770~s mode as
the highest amplitude mode in the Fourier transform, as in 1990 and 1994.
We do not have data to fill in the gap between September
1996 and April 1997 to see how the amplitude changed, but even by
August 19th, the modes at $k=15$ and 18 were already starting to
appear. The time scale
which the star took to change from its normal multi--mode state to
a single mode pulsator was very short, about one day.
The reverse transition started one week after the event.
An estimate of the total energy observed in pulsations is best
obtained by measuring the peak-to-peak amplitudes in the light
curves directly, instead of adding the total power from all the modes.
For the largest amplitude run in 1996, an-0034, observed with the
82" telescope at McDonald, we estimate a peak-to-peak semi-amplitude of
220~mma.
For comparison, the measured Fourier amplitude for the $k=8$ mode for that run
is 170~mma.
For two runs at the same telescope in 2000, we obtain a peak-to-peak
semi-amplitude of 120~mma. Again for comparison, the Fourier amplitudes of the
large
modes present are 30~mma, but there are several modes, and many
combination
peaks. As the observed pulsation
energy is quadratic in the amplitudes and the frequencies,
it corresponds to an
increase of around 34\% in the radiated energy by pulsations,
from the amplitudes, plus a factor of 2.8 from the frequency.
Just two days after the ``forte'', the peak-to-peak
amplitude decreased
by a factor of 5, but during our observations a month later, it had already
increased to its pre-``forte'' value.
It is important to notice that the observed amplitude is not directly
a measurement of the physical amplitude, as there are several factors
that typically depend on $\ell$, including: geometrical
cancellation, inclination effects,
kinetic energies associated with the oscillatory mass motions,
together with a term that depends on the frequency of pulsation
squared.
If we assume that the inclination angle of the pulsation axis to
our line of sight does not change, and that the $\ell$ values of the
dominant modes do not change, then it must be the $\ell$ distribution of the
combination frequencies that changes and produces a difference in the
peak-to-peak variations in the light curve, {\it if the total energy
is conserved}.  This is plausible, as
relatively small variations in the amplitudes of the dominant periods
can dramatically change the amplitudes of the linear combination
frequencies, but not necessary.

\begin{table}
\begin{center}
\begin{tabular}{l|l|c|r|r}
\hline
Telescope & Run & Date (UT) & Time(UT) & Length (s) \\
\hline
\hline

PdM 2m         &   gv-0480 &  1996 Sep 10 & 20:29:01 &  5670 \\
Suhora 60cm      &   suh-62  &  1996 Sep 11 & 18:11:00 & 10790 \\
PdM 2m         &   gv-0484 &  1996 Sep 14 & 21:22:02 &  2330 \\
Suhora 60cm      &   suh-63  &  1996 Sep 18 & 18:45:00 & 15860 \\
Suhora 60cm      &   suh-65  &  1996 Sep 19 & 18:06:20 & 13380 \\
\hline
McD 2.1m         &  an-0061 & 1997 Apr 1 & 06:54:20 & 415 \\
McD 2.1m         &  an-0066 & 1997 Apr 7 & 06:52:50 & 1763 \\
\hline
\hline
\end{tabular}
\end{center}
\caption[Journal of Observations for September 1996 and April 1997]
{Journal of Observation for September, 1996 and April, 1997. September,
1996  data were taken by our
Whole Earth Telescope collaborators during a WET run whose primary target
was not GD~358. PdM stands for Pic du Midi in France, and Suhora is for
Mt. Suhora in Poland. 
The 1997 data were all taken at McDonald Observatory in Texas.
\label{journal2}}
\end{table}

\section{Main Periodicities in the 2000 WET Data}

\subsection{Assumptions and Ground Rules Used in this Work}

We observed GD~358 as the primary target in May-June of 2000 to
provide another ``snapshot'' of the behavior of GD~358 with minimal
alias problems. For the period May 23rd to 
June 8th, this run provided coverage ($\sim 80${\%})
that was intermediate between the 1994 run
(86{\%} coverage) and the 1990 run (with 69{\%} coverage).
The 2000 WET run had several objectives: 1) look for additional
modes besides the known $\ell =1$ $k=8$ through 18 modes;
2) investigate the multiplet splitting structure of the pulsation modes;
3) look for amplitude changes of the known modes;
4) determine the structure of the ``combination peaks'', including
the maximum order seen; and 5) provide simultaneous observations for
HST time resolved spectroscopy.

Before we can start interpreting the peaks in the FT, we need to
select an amplitude limit for what constitutes a ``real'' peak
versus a ``noise'' peak.
Kepler (1993) and Schwarzenberg-Czerny (1991, 1999), following
Scargle (1982), demonstrated that non-equally spaced data sets
and multiperiodic light curves, as
all the Whole Earth Telescope data sets are, do not have a normal
noise distribution, because the residuals are correlated.
The probability that a peak
in the Fourier transform has a $1/1000$ chance of being due
to noise, not a real signal, for our large frequency range of
interest,{\footnote{corresponding to a number of multiple trials
larger than the number of data points}}
requires at least peaks above $4\langle {\mathrm{Amp}}\rangle$,
where the
average amplitude  $\langle {\mathrm{Amp}}\rangle$ is the square root
of the power average (see also Breger et al. 1993 and
Kuschnig et al. 1997 for a similar estimative).

\begin{table}
\begin{center}
\caption{Average amplitude of datasets, from 1000 to 3000~$\mu$Hz.
\label{avamp}}
\begin{tabular}{|l|c|c|}
Year&${\mathrm{BCT_{start}}}$&$\langle {\mathrm{Amp}}\rangle$\\
&(days)&(mma)\\
1990&244\,8031.771867&0.62\\
1994&244\,9475.001705&0.58\\
1996&245\,0307.617884&1.44\\
2000&245\,1702.402508&0.29\\
\end{tabular}
\end{center}
\end{table}

Table~\ref{avamp}
shows that the noise,
represented by $\langle {\mathrm{Amp}}\rangle$, for the 2000 data set is
the smallest to date, allowing us to detect smaller amplitude
peaks. Several peaks in the multi-frequency fits are below the
$4\langle {\mathrm{Amp}}\rangle$ limit and therefore should be
considered only as upper limits to the components.

The present mode identification follows that of the 1990 data set,
published by Winget et al. (1994).
They represent the pulsations in terms of spherical harmonics
$Y_{\ell,m}$, with each eigenmode described by three quantum
numbers: the radial overtone number $k$, the degree $\ell$,
also called the angular momentum quantum number, and the
azimuthal number $m$, with $2\ell+1$ possible values, from
$-\ell$ to $+\ell$.
For a perfectly spherical star, all
$(2\ell+1)$ eigenmodes with the same values of $k$ and $\ell$ should
have the same frequency, but rotation
causes each eigenmode to have a frequency also dependent on $m$.
Magnetic fields also lift the $m$ degeneracy.
The assigned radial order $k$ value
are the outcome of a comparison with model calculations presented
in Bradley \& Winget (1994), and are consistent with the observed mass and
parallax, as discussed in their paper. Vuille et al. (2000) determinations
followed the above ones.
In the upper part of Fig.(\ref{dft2000}), we placed a mark for
equally spaced periods (correct in the asymptotic limit),
using the 38.9~s spacing derived by Vuille et al., starting with
the $k=17$ mode. The observed period spacings in the FT are
very close to equal, consistent with previous observations.

\subsection{Nonlinear Least Squares Results from 1990, 1994, and 1996}

For a more self-consistent comparison, we took the data from
the 1990, 1994 WET runs and the August 1996 run and derived the periods
of the dominant modes via a nonlinear least squares fit.
In Tables~\ref{multif1990}, \ref{multif1994} and \ref{multif1996}
we present the results of a non-linear simultaneous
least squares fit of 23 to 29 sinusoids, representing the main periodicities,
to the 1990, 1994 and 1996 data sets.
We use the nomenclature $k^a$, for example $15^{-}$, to
represent a subcomponent with $m=-1$ of the $k=15$ mode in these tables.
The difference in the frequencies reported in this paper compared to the
previous ones is due to our use of the simultaneous non-linear least-squares
frequency fitting
rather than using the Fourier Transform frequencies.

\begin{table}
\caption{Our multisinusoidal fit to the main periodicities in 1990.\label{multif1990}}
  \begin{tabular} {|l|c|r|r|}
  \hline
  k &Frequency  &   Amplitude &  $T_{max}$ \\
  &$(\mu Hz)$&    $(mma)$  &  $(s)$     \\
  \hline \hline
18&$  1233.408 \pm    0.017$ &  $  5.05 \pm   0.13$ & $731 \pm   7$ \\
$17^+$&$  1291.282 \pm    0.018$ & $  5.04 \pm   0.14$ & $395 \pm   6$ \\
17&$  1297.590 \pm    0.006$ &  $ 14.60 \pm   0.14$ & $ 24 \pm   2$ \\
$17^-$&$  1303.994 \pm    0.019$ &  $  4.71 \pm   0.14$ & $411 \pm   7$ \\
$16^+$&$  1355.664 \pm    0.106$ &  $  0.87 \pm   0.14$ & $471 \pm  37$ \\
16&$  1361.709 \pm    0.040$ &  $  2.21 \pm   0.14$ & $ 18 \pm  14$ \\
$16^-$&$  1368.568 \pm    0.031$ & $  2.96 \pm   0.14$ & $627 \pm  11$ \\
$15^+$&$  1420.932 \pm    0.010$ & $  9.32 \pm   0.14$ & $416 \pm   3$ \\
15&$  1427.402 \pm    0.005$ &  $ 19.24 \pm   0.14$ & $425 \pm   2$ \\
$15^-$&$  1433.853 \pm    0.011$ & $  7.90 \pm   0.14$ & $ 88 \pm   4$ \\
$14^+$&$  1513.023 \pm    0.017$ & $  5.23 \pm   0.14$ & $123 \pm   5$ \\
14&$  1518.991 \pm    0.009$ &  $  9.71 \pm   0.14$ & $270 \pm   3$ \\
$14^-$&$  1525.873 \pm    0.016$ & $  5.35 \pm   0.13$ & $459 \pm   5$ \\
$13^+$&$  1611.671 \pm    0.016$ & $  5.70 \pm   0.14$ & $140 \pm   5$ \\
13&$  1617.297 \pm    0.017$ & $  5.28 \pm   0.14$ & $508 \pm   5$ \\
$13^-$&$  1623.644 \pm    0.019$ & $  4.70 \pm   0.14$ & $407 \pm   5$ \\
12&$  1733.850 \pm    0.163$ & $  0.53 \pm   0.13$ & $474 \pm  44$ \\
$11^+$&$  1840.022 \pm    0.136$ &  $  0.65 \pm   0.14$ & $ 41 \pm  35$ \\
11&$  1846.247 \pm    0.135$ & $  0.66 \pm   0.14$ & $504 \pm  34$ \\
$11^-$&$  1852.099 \pm    0.093$ &  $  0.94 \pm   0.14$ & $132 \pm  24$ \\
$10^+$&$  1994.240 \pm    1.071$ &  $   \leq 0.14$ &  \\
10&$  1998.919 \pm    0.060$ &  $  1.50 \pm   0.14$ & $ 25 \pm  14$ \\
$10^-$&$  2007.992 \pm    0.117$ &  $  0.84 \pm   0.14$ & $  7 \pm  26$ \\
$9^+$&$  2150.430 \pm    0.048$ &  $  1.932 \pm   0.13$ & $174 \pm  10$ \\
9&$  2154.052 \pm    0.020$ &  $  4.59 \pm   0.13$ & $336 \pm   4$ \\
$9^-$&$  2157.834 \pm    0.032$ &  $  2.81 \pm   0.13$ & $400 \pm   7$ \\
$8^+$&$  2358.975 \pm    0.016$ &  $  5.68 \pm   0.13$ & $120 \pm   3$ \\
8&$  2362.588 \pm    0.016$ &  $  5.77 \pm   0.13$ & $422 \pm   3$ \\
$8^-$&$  2366.418 \pm    0.017$ & $  5.34 \pm   0.13$ & $268 \pm   3$ \\
  \hline
  \end{tabular}
\end{table}

  \begin{table}
\caption{Main periodicities in 1994.\label{multif1994}}
  \begin{tabular} {|l|c|r|r|}
  \hline
  k&Frequency  & Amplitude &  $T_{max}$ \\
  &$(\mu Hz)$&  $(mma)$  &  $(s)$     \\
  \hline \hline
$18^+$&$1228.712 \pm    0.022$ & $  2.77 \pm   0.13$ & $252.6 \pm  12.0$ \\
18&$1235.493 \pm    0.005$ & $ 12.94 \pm   0.13$ & $170.9 \pm   2.6$ \\
$18^-$&$1242.364 \pm    0.016$ & $  3.66 \pm   0.13$ & $ 62.4 \pm   9.0$ \\
$17^+$&$1291.093 \pm    0.010$ & $  6.17 \pm   0.13$ & $250.3 \pm   5.1$ \\
17&$1297.741 \pm    0.003$ & $ 22.11 \pm   0.13$ & $ 37.7 \pm   1.4$ \\
$17^-$&$1304.459 \pm    0.010$ & $  6.25 \pm   0.13$ & $615.0 \pm   5.0$ \\
$16^+$&$1355.388 \pm    0.035$ & $  1.70 \pm   0.13$ & $167.0 \pm  17.6$ \\
16&$1362.298 \pm    0.060$ & $  <0.89$& \\
$16^-$&$1368.541 \pm    0.031$ & $  1.92 \pm   0.13$ & $322.3 \pm  15.5$ \\
$15^+$&$1419.650 \pm    0.003$ & $ 18.37 \pm   0.13$ & $ 46.4 \pm   1.6$ \\
15&$1426.408 \pm    0.004$ & $ 15.55 \pm   0.13$ & $239.7 \pm   1.8$ \\
$15^a$&$1430.879 \pm    0.006$ & $ 10.61 \pm   0.13$ & $187.9 \pm   2.7$ \\
$15^-$&$1433.169 \pm    0.014$ & $  4.46 \pm   0.13$ & $104.1 \pm   6.5$ \\
14&$1519.903 \pm    0.028$ & $  1.09 \pm   0.13$ & $485.2 \pm   20.0$ \\
$13^+$&$1611.357 \pm    0.012$ & $  5.02 \pm   0.13$ & $466.0 \pm   5.0$ \\
13&$1617.474 \pm    0.009$ & $  3.46 \pm   0.13$ & $183.3 \pm   1.1$ \\
$13^-$&$1624.568 \pm    0.010$ & $  6.07 \pm   0.13$ & $101.0 \pm   4.1$ \\
12&$1746.766 \pm    0.064$ & $  0.93 \pm   0.13$ & $414.2 \pm  25.0$ \\
11&$1863.004 \pm    0.184$ & $  <0.71$ &\\
10&$2027.325 \pm    0.457$ & $  <0.46$ &\\
$9^+$&$2150.504 \pm    0.019$ & $  3.15 \pm   0.13$ & $346.7 \pm   6.1$ \\
9&$2154.124 \pm    0.013$ & $  4.76 \pm   0.13$ & $ 12.3 \pm   4.0$ \\
$9^-$&$2157.841 \pm    0.022$ & $  2.69 \pm   0.13$ & $144.2 \pm   7.1$ \\
$8^+$&$2358.883 \pm    0.013$ & $  4.50 \pm   0.13$ & $398.2 \pm   3.8$ \\
8&$2362.636 \pm    0.006$ & $  9.25 \pm   0.13$ & $274.6 \pm   1.8$ \\
$8^-$&$2366.508 \pm    0.007$ & $  4.22 \pm   0.13$ & $169.7 \pm   3.8$ \\
  \hline
  \end{tabular}
  \end{table}

\begin{table}
\caption{Main modes in 1996.\label{multif1996}}
  \begin{tabular} {|l|c|r|r|}
  \hline
  k&Frequency  & Amplitude &  $T_{max}$ \\
  &$(\mu Hz)$&  $(mma)$  &  $(s)$     \\
  \hline \hline
19&$1172.66 \pm    0.15$ & $  2.5 \pm   0.6$ & $ 17.4 \pm  47.5$ \\
18&$1253.65 \pm    1.01$ & $  < 2.1$ &\\
$17^+$&$1291.13 \pm    0.11$ & $  4.3 \pm   0.6$ & $653.6 \pm  28.1$ \\
$17^0$&$1295.38 \pm    0.17$ & $  2.6 \pm   0.6$ & $139.7 \pm  43.6$ \\
$17^-$&$1304.68 \pm    0.11$ & $  4.8 \pm   0.6$ & $346.3 \pm  25.7$ \\
$16^+$&$1355.21 \pm    2.02$ & $  < 1.9$ & \\
$16^0$&$1362.55 \pm    0.15$ & $  2.7 \pm   0.6$ & $172.2 \pm  41.2$ \\
$16^-$&$1379.64 \pm    0.16$ & $  2.7 \pm   0.6$ & $378.6 \pm  41.0$ \\
$15^0$&$1427.47 \pm    0.92$ & $  < 2.2$&  \\
$15^-$&$1434.36 \pm    0.18$ & $  2.2 \pm   0.6$ & $154.5 \pm  47.1$ \\
14&$1520.58 \pm    0.18$ & $  2.0 \pm   0.6$ & $398.3 \pm  46.6$ \\
$13^+$&$1611.60 \pm    0.18$ & $  2.1 \pm   0.6$ & $461.8 \pm  42.7$ \\
$13^0$&$1617.51 \pm    0.35$ & $  1.1 \pm   0.6$ & $183.5 \pm  79.9$ \\
$13^-$&$1619.63 \pm    0.84$ & $  < 2.2$ &  \\
12&$1736.10 \pm    0.34$ & $  1.1 \pm   0.6$ & $323.6 \pm  75.2$ \\
11&$1862.93 \pm    0.39$ & $  0.9 \pm   0.6$ & $ 58.8 \pm  78.7$ \\
10&$2027.41 \pm    0.26$ & $  1.4 \pm   0.6$ & $471.9 \pm  47.9$ \\
$9^+$&$2149.97 \pm    0.07$ & $  5.6 \pm   0.6$ & $ 69.8 \pm  11.8$ \\
$9^0$&$2153.84 \pm    0.05$ & $  7.6 \pm   0.6$ & $155.2 \pm   8.5$ \\
$9^-$&$2157.89 \pm    0.04$ & $  9.1 \pm   0.6$ & $426.7 \pm   7.2$ \\
$8^+$&$2358.63 \pm    0.03$ & $ 12.6 \pm   0.6$ & $125.3 \pm   4.7$ \\
$8^0$&$2362.50 \pm    0.02$ & $ 23.2 \pm   0.6$ & $104.8 \pm   2.5$ \\
$8^-$&$2365.98 \pm    0.02$ & $ 22.2 \pm   0.6$ & $142.9 \pm   2.6$ \\
  \hline
  \end{tabular}
\end{table}

We note that both the Fourier analysis and multi-sinusoidal fit
assume the signal is composed of sinusoids with constant amplitudes,
which is clearly violated in the 1996 data set.
The changing amplitudes introduce spurious peaks in the Fourier transform.
This will not affect the frequency of the modes, but the inferred
amplitude will be a poor match to the (non-sinusoidal) light curve amplitude.

In Table~\ref{avamp} we present the
average amplitude of the data sets, from 1000 to 3000~$\mu$Hz,
after the main periodicities, all above $4\langle {\mathrm{Amp}}\rangle$,
have been subtracted. For the 2000 data set, the initial
$\langle {\mathrm{Amp}}\rangle$ for the frequency range from 0
to 10\,000~$\mu$Hz, is 0.69~mma. For the high frequency range above
3000~$\mu$Hz, $\langle {\mathrm{Amp}}\rangle\simeq 0.2$~mma.

\subsection{Mode Analysis of the 2000 WET Data}

To provide the most accurate frequencies possible, we rely on a non-linear
least squares fit of sinusoidal modes with guesses to the observed periods,
since these better take into account contamination or slight frequency
shifts due to aliasing.
In Table~\ref{multif2000} we present the results of a simultaneous
non-linear least squares fit of 29 sinusoids, representing the main
periodicities of the 2000 data set, simultaneously.
All the phases have been measured with respect to the
barycentric Julian coordinated date
BCT 2\,451\,702.402\,508.

\begin{table} 
\caption{Main modes in 2000.\label{multif2000}}
  \begin{tabular} {|l|c|r|r|r|}
  \hline
k& Frequency  & Period\hfil    &  Amplitude\hfil &  $T_{max}$\hfil \\
& $(\mu Hz)$&  $(sec)$  &   $(mma)$\hfil  &  $(s)$\hfil     \\
  \hline \hline
20&$  1110.960 \pm    0.017$ & $ 900.122 \pm    0.014$ & $  2.04 \pm   0.08$ &
$870.19 \pm   8.61$ \\
19&$  1172.982 \pm    0.013$ & $ 852.528 \pm    0.009$ & $  2.74 \pm   0.08$ &
$164.86 \pm   6.07$ \\
$\ell=2$&$  1255.400 \pm    0.002$ & $ 796.556 \pm    0.002$ & $ 14.86 \pm  
0.08$ & $747.02 \pm   1.05$ \\
18&$  1233.595 \pm    0.018$ & $ 810.639 \pm    0.012$ & $  1.96 \pm   0.08$ &
$354.91 \pm   8.11$ \\
$17^+$&$  1294.284 \pm    0.094$ & $ 772.628 \pm    0.056$ & $  0.38 \pm  
0.082$ & $579.63 \pm  39.54$ \\
17&$  1296.599 \pm    0.001$ & $ 771.248 \pm    0.001$ & $ 29.16 \pm   0.08$ &
$247.81 \pm   0.52$ \\
$17^-$&$  1301.653 \pm    0.053$ & $ 768.254 \pm    0.031$ & $  0.68 
\pm   0.08$
& $314.52 \pm  22.52$ \\
16&$  1362.238 \pm    0.159$ & $ 734.086 \pm    0.086$ & $  0.42 \pm   0.12$ &
$263.36 \pm  63.93$ \\
$16^-$&$  1378.806 \pm    0.007$ & $ 725.265 \pm    0.004$ & $  5.35 
\pm   0.08$
& $514.70 \pm   2.66$ \\
$15^+$&$  1420.095 \pm    0.001$ & $ 704.178 \pm    0.001$ & $ 29.69 \pm  
0.08$ & $418.14 \pm   0.49$ \\
15&$  1428.090 \pm    0.052$ & $ 700.236 \pm    0.025$ & $  0.70 \pm   0.08$ &
$217.29 \pm  19.78$ \\
$15^-$&$  1432.211 \pm    0.036$ & $ 698.221 \pm    0.018$ & $  1.08 \pm  
0.089$ & $355.58 \pm  13.30$ \\
14&$  1519.811 \pm    0.134$ & $ 657.977 \pm    0.058$ & $  0.266 
\pm   0.08$ &
$301.55 \pm  48.28$ \\
$13^+$&$  1611.084 \pm    0.116$ & $ 620.700 \pm    0.045$ & $  0.31 
\pm   0.08$
& $448.66 \pm  39.81$ \\
13&$  1617.633 \pm    0.174$ & $ 618.187 \pm    0.066$ & $  0.21 \pm   0.08$ &
$448.19 \pm  58.97$ \\
$13^-$&$  1625.170 \pm    0.235$ & $ 615.320 \pm    0.089$ & $  0.15 
\pm   0.08$
& $299.24 \pm  79.32$ \\
12&$  1736.277 \pm    0.034$ & $ 575.945 \pm    0.011$ & $  1.04 \pm   0.08$ &
$230.37 \pm  10.79$ \\
11&$  1862.871 \pm    0.042$ & $ 536.806 \pm    0.012$ & $  0.84 \pm   0.08$ &
$503.66 \pm  12.43$ \\
10&$  2027.008 \pm    0.028$ & $ 493.338 \pm    0.007$ & $  1.29 \pm   0.08$ &
$350.14 \pm   7.49$ \\
$9^+$&$  2150.462 \pm    0.012$ & $ 465.016 \pm    0.003$ & $  2.96 
\pm   0.08$
& $ 35.49 \pm   3.09$ \\
9&$  2154.021 \pm    0.007$ & $ 464.248 \pm    0.001$ & $  5.34 \pm   0.08$ &
$252.38 \pm   1.71$ \\
$9^-$&$  2157.731 \pm    0.014$ & $ 463.450 \pm    0.003$ & $  2.57 
\pm   0.08$
& $174.68 \pm   3.53$ \\
$8^+$&$  2359.119 \pm    0.006$ & $ 423.887 \pm    0.001$ & $  5.57 
\pm   0.08$
& $166.60 \pm   1.49$ \\
8&$  2362.948 \pm    0.094$ & $ 423.200 \pm    0.017$ & $  0.38 \pm 
0.08$ & $
81.50 \pm  21.93$ \\
$8^-$&$  2366.266 \pm    0.006$ & $ 422.607 \pm    0.001$ & $  5.79 
\pm   0.08$
& $418.36 \pm   1.44$ \\
$2\times 18$&$  2510.761 \pm    0.021$ & $ 398.286 \pm    0.003$ & $ 
1.70 \pm  
0.08$ & $334.47 \pm   4.58$ \\
$2\times 17$&$  2593.208 \pm    0.004$ & $ 385.623 \pm    0.001$ & $ 
7.82 \pm  
0.08$ & $249.57 \pm   0.96$ \\
7&$  2675.487 \pm    0.004$ & $ 373.764 \pm    0.001$ & $  8.49 \pm   0.08$ &
$193.30 \pm   0.86$ \\
$2\times 15$&$  2840.195 \pm    0.008$ & $ 352.089 \pm    0.001$ & $ 
4.29 \pm  
0.08$ & $ 47.50 \pm   1.60$ \\
  \hline
  \end{tabular}
\end{table}

Armed with the new frequencies in Table~\ref{multif2000}, we comment
on regions of particular interest in the FT.
First, we identify several newly detected modes at
$P=373.76$~s, $f=2675.49~\mu$Hz, amp=8.43~mma;
$P=852.52$~s, $f=1172.99~\mu$Hz, amp=2.74~mma; and
$P=900.13$~s, $f=1110.95~\mu$Hz, amp=2.03~mma.
Based on the mode assignments of Bradley {\&} Winget (1994) we identify
these modes as $k=7$, 19, and 20.
The mode identification is based on the proximity of
the detected modes with those predicted by the models, or even the
asymptotical period spacings, but also because of resonant mode coupling,
i.e., a stable mode will be driven to visibility if a coupled mode
falls near its frequency, as it happens for $k=7$, which is very close to
the combination of $k=17$ and $k=16$, and $k=20$, which falls near
the resonance of the $8^-$ and the $\ell =2$ mode at $1255~\mu$Hz
(see next paragraph).
It is important to note that these modes appear in combination peaks
with other modes, as shown in Table~\ref{combination}.
This reinforces our belief that these modes are physical modes, and not
just erroneously identified combination peaks.
We note that Bradley (2002) analyzed single site data taken over several
years, and found peaks at 1172 or $1183~\mu$Hz in
April 1985, May 1986, and June 1992 data, lending additional credence to
the detection of the $k=19$ mode or its alias.

\begin{figure}
\centering
\includegraphics[width=\linewidth]{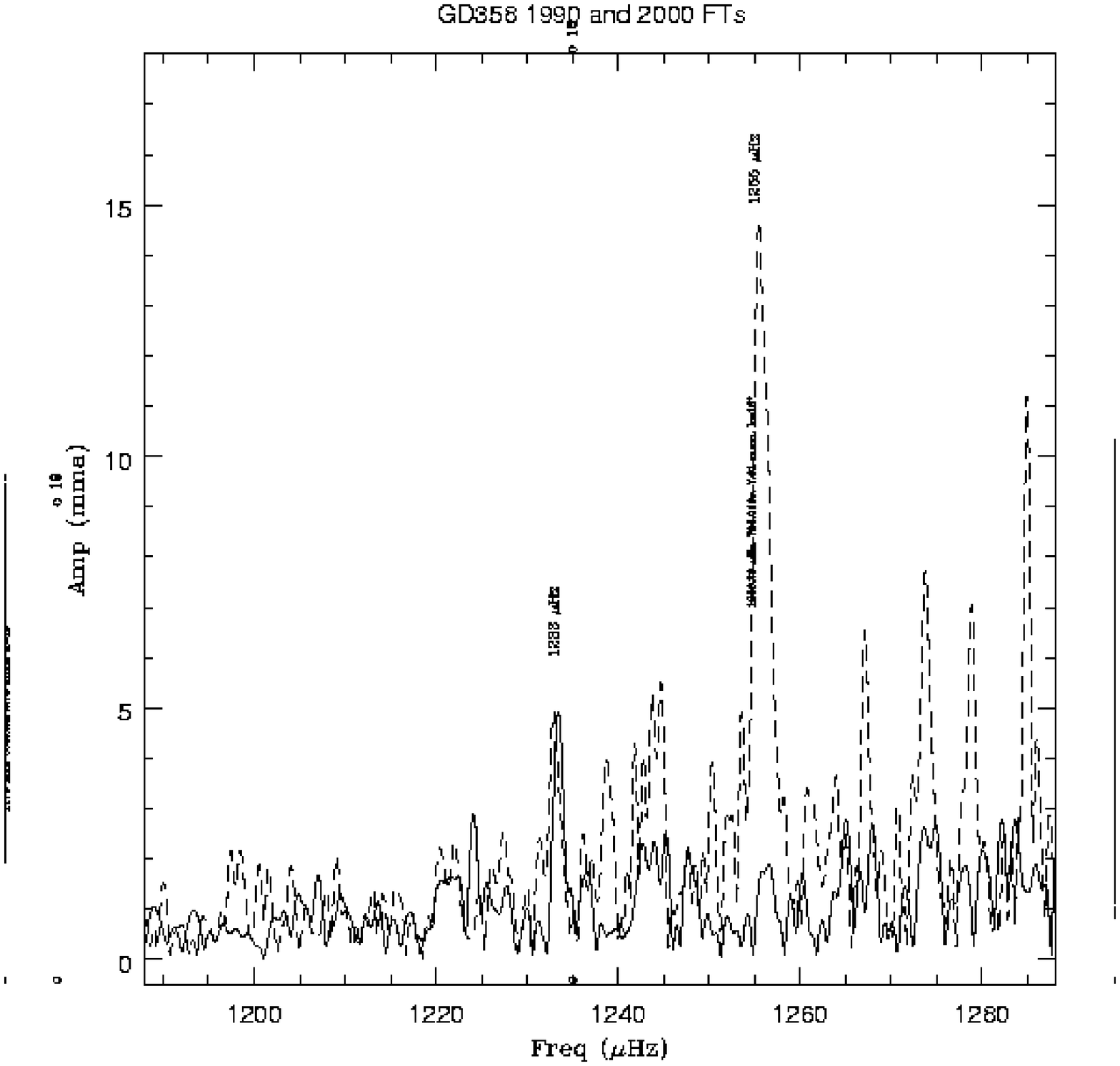}
\caption{Peaks around k=18 in the 1990 (solid line) and 2000 (dashed line)
transforms}
\label{k18}
\end{figure}

The first previously known region of interest surrounds the $k=18$ mode,
which lies
near $1233~\mu$Hz, according to previous observations.
In the 2000 data, the largest amplitude peak in this region lies at
$1255~\mu$Hz, which is over $20~\mu$Hz from the previous location.
Given that other modes (especially the one at $k=17$) has shifted by
less than $4~\mu$Hz, we are inclined to rule out the possibility that
the $k=18$ mode shifted by $20~\mu$Hz.
One possible solution is offered by seismological models of GD~358,
which predict an $\ell =2$ mode near $1255~\mu$Hz.
For example, the best ML2 fit to the 1990 data (from
Metcalfe, Salaris \& Winget 2002, Table 3), has an $\ell=2$ mode
at $1252.6~\mu$Hz (P=798.3~s).
This would also be consistent with the larger number of subcomponents
detected, although they may be caused only by amplitude changes during
the observations. Fig (\ref{k18}) shows the region of the  $k=18$ mode
in the FT for the 1990 data set (solid) and the 2000 data set (dashed);
it is consistent with the $k=18$ mode being the $1233~\mu$Hz for both data
sets, and they even have similar amplitudes.
While we avoided having to provide an explanation for why only
the $k=18$ mode would shift by $20~\mu$Hz, we have introduced another
problem, as geometrical cancellation for an $\ell=2$ mode
introduces a factor of 0.26 in relation to unity for an
$\ell=1$ mode. Thus, the identification of the $1255~\mu$Hz
as an $\ell=2$ mode, which has a measured amplitude of 14.86~mma,
implies a physical amplitude higher than that of the highest
amplitude $\ell=1$ mode, around 30~mma.

Kotak et al. (2002), analyzing time-resolved spectra obtained at the
Keck in 1998, show the velocity variations of the $k=18$ mode
at $1233~\mu$Hz is similar to those for the $k=15$ and $k=17$ modes,
concluding all modes are $\ell=1$. They did not detect a mode
at $1255~\mu$Hz.

\begin{figure}
\centering
\includegraphics[width=\linewidth]{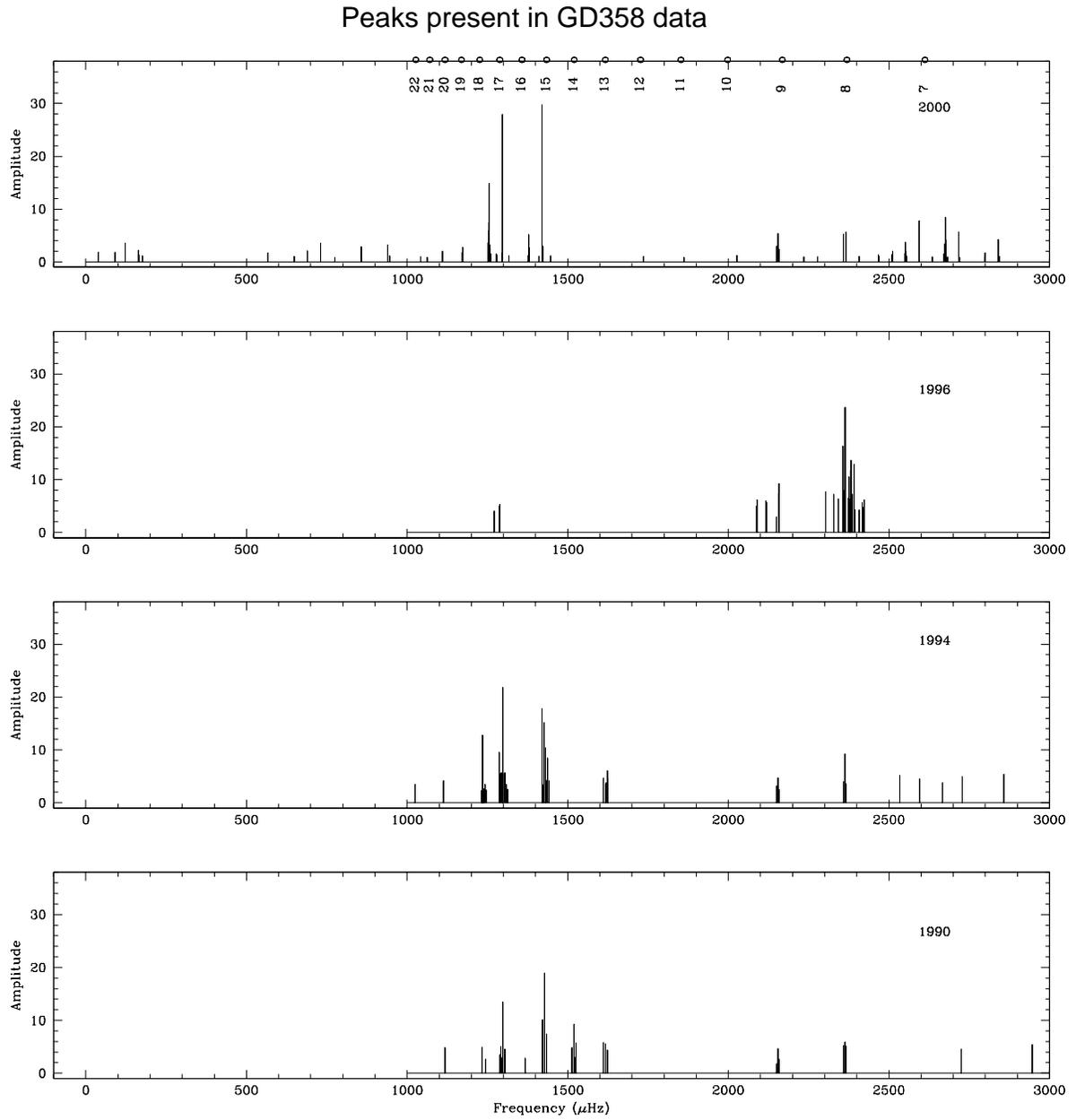}
\caption{Results of pre-whitening for the 1990, 1994, August 1996,
and 2000 data sets}
\label{pkh}
\end{figure}

In Fig. \ref{pkh} we show the pre-whitened results; pre-whitening
was done by subtracting from the observed light curve a synthetic light
curve constructed with a single sinusoid with frequency, amplitude
and phase that minimizes the Fourier spectrum at the frequency of the
highest peak. A new Fourier spectrum is calculated and the next dominant
frequency is subtracted, repeating the procedure until no significant
power is left.
It is important to notice that with  pre-whitening, the order of
subtraction matters. As an example, in the 2000 data set, if we
subtract the largest peak in the region of the $k=18$ mode, at
1255.41~$\mu$Hz, followed by the next highest peak at 1256.26~$\mu$Hz
and the next at 1254.44~$\mu$Hz, we are left with a peak at only
1.3~mma at 1232.76~$\mu$Hz. But if instead we subtract only the
1255.41~$\mu$Hz
followed by the peak left at 1233.24~$\mu$Hz, its amplitude is
around 3.1~mma, i.e., larger. Pre-whitening assumes the frequencies are
independent in the observed, finite, data set.
If they were, the order of subtraction would not affect the result.
Because the order of subtraction matters, the basic assumption
of pre-whitening does not apply.
We attempt to minimize this effect by noting that the frequencies
change less than the amplitudes, and use the FT frequencies in a
simultaneous non-linear least squares fit of all the eigenmode
frequencies.
But even the simultaneous
non-linear least squares fit uses the values of the Fourier
transform as starting points, and could converge to a local
minimum of the variance instead of the global minimum.

The modes with periods between 770 and 518~s ($k=17$ through 13) are
present in the 2000 data, though with different amplitudes than in
previous years.
Another striking feature of the peaks in 2000 is that one multiplet
member of each mode has by far the largest amplitude, so that without
data from previous WET runs, we would not know that the modes are
rotationally split.
The frequencies of these modes are stable to about $1~\mu$Hz
or less with the exception of the $16^-$ mode, where the frequency
jumped from about $1368.5~\mu$Hz in 1990 and 1994 to about $1379~\mu$Hz
in 1996 and 2000 (see Fig. 11).
Most of these frequency changes are larger than the formal frequency
uncertainty from a given run (typically less than $0.05~\mu$Hz), so
there is some process in GD~358 that causes the mode frequencies to
``wobble'' from  one run to the next.
We speculate that this may be related to non-linear mode coupling
effects.
Whatever the origin of the frequency shifts, it renders these modes
useless for studying evolutionary timescales through rates of period
change.

The $k=12$ through $10$ modes deserve separate mention because their
amplitudes are always small; between 1990 and 2000, the largest amplitude
peak was only $1.6$~mma.
The small amplitudes can make accurate frequency determinations
difficult, and all three modes have frequency shifts of 13 to $33~\mu$Hz
between the largest amplitude peaks in a given mode.
The $k=10$ mode shows the largest change with the 1990 data showing
the largest peaks at $1998.7~\mu$Hz and $2008.2~\mu$Hz, while the 2000
data has one peak dominating the region at $2027.0~\mu$Hz.
An examination of the data in Bradley (2002) shows that the $k=12$ mode
seems to consistently show a peak near $1733$ to $1736~\mu$Hz, and that
only the 1994 data has the peak shifted to $1746.8~\mu$Hz, suggesting
that 1994 data may have found an alias peak or that the $1736~\mu$Hz
mode could be the $m=+1$ member and the $1746.8~\mu$Hz mode is the
$m=-1$ member. The data in Bradley (2002) do not show convincing
evidence for the $k=11$ or 10 modes, so we cannot say anything else
about them.

It is interesting to note that the $k=8$ and
$k=9$ modes are always seen as a triplet,
with 3.58~$\mu$Hz separation for $k=9$, even in the 1996 data set.
Our measured spacings are 3.54 and 3.69~$\mu$Hz, from $m=-1$ to $m=0$ and
$m=0$ to $m=1$.
The $k=8$ mode
in 2000 shows an m=0 component below our statistical
detection limit (A=0.41~mma, when the local
$\langle {\mathrm{Amp}} \rangle=0.29$~mma),
but the $m=1$ and $m=-1$ modes
remain separated by $2\times 3.58~\mu$Hz. All the higher k modes
are seen as singlets in the 2000 data set.
We also note that the $k=8$ and 9 modes have by far the most
stable frequencies.
The frequencies are always the same to within $0.3~\mu$Hz, and in
some cases better than $0.1~\mu$Hz.
However, the frequency shifts are large enough to mask any possible
signs of evolutionary period change, as Fig.~12 shows.
Thus, we are forced to conclude that GD~358 is not a stable enough
``clock'' to discern evolutionary rates of period change.

\section{Multiplet Splittings}

As pointed out by Winget et al. (1994) and Vuille et al. (2000),
the observed triplets in the 1900 and 1994 data sets had
splittings ranging from 6.5~$\mu$Hz from the ``external'' modes
(such as $k=17$) to 3.6~$\mu$Hz for the ``internal'' modes
$k=8$ and 9.
Winget et al. interpreted these splittings to be the result of
radial differential rotation, and Kawaler et~al. (1999) examined
this interpretation in more detail.
An examination of the frequencies found in the 2000 data set,
shown in Table~\ref{multif2000}, shows that the multiplet structure
is much harder to discern, since the $k=10$ through 20 modes typically
have only one multiplet member with a large amplitude.
The obvious multiplet members have frequencies that agree with the
1990 data, except for the $16^-$ mode, where there is a $+10.234~\mu$Hz
shift in the 2000 data.

Note that the $k=10$ mode identified at $2027~\mu Hz$ is different than the
$1994~\mu Hz$ identified by Winget et al. (1994) in the 1990 data.
However, the
peak they identified is not the highest peak in that region of the Fourier
transform (see Fig. \ref{k10}).
Our analysis of the 1990 data has statistically significant $k=10$
peaks close to $1999~\mu$Hz and 2008~$\mu$Hz.

The only modes with obvious multiplet structure
are the $k=9$ mode, which still shows an obvious
3.6~$\mu$Hz split triplet, and
the $k=8$ mode, which shows two peaks that are
consistent with $2\times 3.6~\mu$Hz separation.

In Table~\ref{combination} we have a peak 3.3~$\mu$Hz from the $k=15$, $m=1$
mode that we have not seen before; we call it the $15^b$ mode.
We are not certain whether this is another member of the $k=15$ multiplet
(analogous to the $15^a$ mode in the 1994 WET data) or if it is
something else.
The 1994 data set also presented a large peak 4.4~$\mu$Hz
from $k=15$, $m=0$, which we call the $15^a$ mode, in addition to the
$m=\pm 1$ components. We have not seen this $15^a$ mode in any other
data set other than the 1994 WET run.
The identity of the ``extra subcomponents'' remains an unsolved mystery.

\section{Linear Combinations}

Winget et al. (1994) and Vuille et al. (2000) show that most of the
periodicities are in fact linear combination peaks of the main
peaks (eigenmodes).
Combination peaks are what we call peaks in the FT whose frequencies
are equal to the sum or difference of two (or more) the the $\ell=1$ or 2
mode frequencies.
The criteria for selection of the combination peaks was that the
frequency difference between the combination peak and the sum of the
``parent mode'' frequencies
must be smaller than our resolution, which is typically around 1~$\mu$Hz.
The last column of Table~\ref{combination} shows the frequency difference.

****
TABLE~\ref{combination}  - Linear combinations (should be here)
***

For example, only 28 of the more than 180 peaks in Winget et~al.
(1994) are $\ell =1$ modes; the rest are combination peaks up to
third order (i.e., three modes are involved).
The $\ell =1$ modes lie in the region 1000 to 2500~$\mu$Hz, and are
identified as modes $k=18$ to 8. In the 1994 data set analyzed
by Vuille et al., combination peaks up to 4th order were detected.
In the 2000 data set we identify combination peaks up to 6th order, and
most if not all remaining peaks are in fact linear combination peaks,
as demonstrated in Table~\ref{combination} and is shown in the pre-whitened
FT of the 2000 data (see Fig. \ref{pk2000}).
Here too, we use the nomenclature $k^a$, for example $15^{-}$, to
represent a subcomponent with $m=-1$ of the $k=15$ mode.

The so-called $\ell=2$ mode at $1255~\mu$Hz,
as well as $k=17$ and $k=15$ modes, have subcomponents, but
probably they are not different $m$ value components,
and are caused, most likely, by amplitude modulation.
We say this because the frequency splittings are drastically
different than in previous data, and for the $\ell =2$ mode, there
are more than 5 possible subcomponent peaks present.
We did not do an exhaustive search for
all of the possible combination peaks up to 6th order in the Fourier
transform, as we only took into account
the peaks that had a probability smaller than 1/1000
of being due to noise, and studied if they could
be explained as combination peaks.

Brickhill's (1992) pulsation--convection interaction model
predicts, and the observations reported by Winget et al. and
Vuille et al. agree, that a combination peak involving
two different modes always has a larger relative amplitude
than a combination involving twice the frequency of a given mode
(also called a harmonic peak).
Wu's (2001) analytical expression leads to a factor of 1/2
difference between a combination peak with two modes versus a
harmonic peak, assuming that the
amplitudes of both eigenmodes are the same.
Vuille et al. claim that the relatively small amplitude of
the $k=13$ mode in 1994 is affected by
destructive beating of the
nonlinear peak ($2\times 15-18$) and that the $k=16$ mode amplitude
is affected by the ($15+18-17$) combination peak.
It is noteworthy that the peak at 1423.62~$\mu$Hz is only 3.52~$\mu$Hz
from $k=15$, so it might be the $15^-$ mode.
However, the previously identified
$15^-$ was 6.7~$\mu$Hz from it, and we consider the $1423.62~\mu$Hz peak
to be either a result of amplitude modulation of the $k=15$ mode or yet
another combination peak.

We note that the wealth of combination peaks and their relative
amplitude offers insight into the amplitude limiting mechanism and
would be worthy of the considerable theoretical and numerical effort
required to understand it.

\section{Model-Fitting with a Genetic Algorithm}

One of the major goals of our observations of GD~358 was to discover
additional modes to help refine our seismological model fits. We were also
interested in how much the globally optimal model parameters would change
due to the slight shifts in the observed periods. With these goals in
mind, we repeated the global model-fitting procedure of Metcalfe, Winget \&
Charbonneau (2001) on several subsets of the new observations.

Our model-fitting method uses the parallel genetic algorithm described by
***
Metcalfe \& Charbonneau (2002) to minimize the root-mean-square (rms)
differences between the observed and calculated periods ($P_k$) {\it and}
period spacings ($\Delta P\equiv P_{k+1}-P_k$) for models with effective
temperatures ($T_{\rm eff}$) between 20,000 and 30,000 K, total stellar
masses ($M_*$) between 0.45 and 0.95 $M_{\odot}$, helium layer masses with
$-\log(M_{\rm He}/M_*)$ between 2.0 and $\sim$7.0, and an internal C/O
profile with a constant oxygen mass fraction ($X_{\rm O}$) out to some
fractional mass ($q$) where it then decreases linearly in mass to zero
oxygen at $0.95~m/M_*$. This technique has been shown to find the globally
optimal set of parameters consistently among the many possible
combinations in the search space, but it requires between $\sim$200 and
4000 times fewer model evaluations than an exhaustive search of the
parameter-space to accomplish this, and has a failure rate $<10^{-5}$.

We attempted to fit the 13 periods and period spacings defined by the
$m=0$ components of the 14 modes identified as $k=7$ to $k=20$ in 
Table~\ref{multif2000}. Because of our uncertainty about the proper 
identification of
$k=18$
(see section 4) we performed fits under two different assumptions: for Fit
$a$ we assumed that the frequency near 1233~$\mu$Hz was $k=18$ (similar to
the frequency identified in 1990), and for Fit $b$ we assumed that the
larger amplitude frequency near 1255~$\mu$Hz was $k=18$. The results of
these two fits led us to prefer the identification for $k=18$ in Fit $a$,
and we included this in an additional fit using only the 11 modes from
$k=8$ to $k=18$, which correspond to those identified in 1990 (Fit $c$).
We performed an additional fit (Fit $d$) that included the same 13 periods
used for Fit $a$, but ignored the period spacings.
The optimal values for the five model parameters, and the root-mean-square
residuals between the observed and computed periods ($\sigma_P$) and
period spacings ($\sigma_{ \Delta P}$) for the four fits are shown in
Table \ref{fits2000}.

\begin{table}
\caption{Optimal Fits to 2000 data.\label{fits2000}}
  \begin{tabular} {|r|r|r|r|r|}
  \hline
Parameter               & Fit $a$ & Fit $b$ & Fit $c$ & Fit $d$ \\
\hline \hline
$T_{\rm eff} (K)$       & 24,300  & 23,500  & 24,500  & 22,700  \\
$M_* (M_{\odot})$       & 0.61    & 0.60    & 0.625   & 0.630   \\
$\log(M_{\rm He}/M_*)$  & -2.79   & -5.13   & -2.58   & -4.07   \\
$X_{\rm O}$             & 0.81    & 0.99    & 0.39    & 0.37    \\
$q (m/M_*)$             & 0.47    & 0.47    & 0.83    & 0.42    \\
$\sigma_P (s)$          & 2.60    & 3.65    & 2.12    & 1.72    \\
$\sigma_{\Delta P} (s)$ & 4.07    & 4.92    & 2.21    & $\cdots$\\
\hline
\end{tabular}
\end{table}

Our preferred solution from Table \ref{fits2000} is Fit $a$, because it
includes our favored identification for the $k=18$ mode and the additional
pulsation periods. The larger $\sigma_{\Delta P}$ in Fit $a$ compared to
Fit $c$ is dominated by the large period spacings between the $k=19$ and
20 modes ($47.6$~s) and the $k=7$ and 8 modes ($49.4$~s). Fit $a$ has a
mass and effective temperature that are essentially the same as the fit of
Bradley {\&} Winget (1994), and are consistent with the spectroscopic
values derived by Beauchamp et al. (1999). The other structural parameters
are otherwise similar to those found by Metcalfe et~al. (2001) 
($T_{\mathrm{eff}}= 22,600$~K, $M_* = 0.650 M_{\odot}$, 
$log(M_{\rm He}/M_* = -2.74$,
$X_{\rm O} = 0.84$, and $q = 0.49$). We caution, however, that the large
values of $\sigma_P$ and $\sigma_{\Delta P}$ for Fit $a$ imply that our
model may not be an adequate representation of the real white dwarf star.
New and unmodeled physical circumstances may have arisen between 1994 and
2000 (e.g. whatever caused the {\it forte} in 1996), which may account for 
the diminished capacity of our simple model to match the observed periods.

\section{Summary of Observational Clues}

Before embarking on our discussion, we recap the highlights of our
observations.
First, the 2000 WET data shows eigenmodes from at least
$k=8$ through $k=19$. We may also have detected the $k=7$ and $k=20$
modes. However, their frequencies are similar to that of unrelated
combination peaks, so their identification is less secure.
For the first time, we have also found an $\ell =2$ mode in the
GD~358 data; it is at $1255.4~\mu$Hz.
Second, we see relatively few multiplet modes for a given $k$, with
the exception of the $k=8$ and 9 modes.
While the multiplet structure of the $\ell =1$ modes is muted, the
combination peaks are enhanced to the point that we see combination
modes up to 6th order. Combined with the previous WET runs, we see
evidence for anticorrelation between the presence of multiplet
structure and combination peaks.
The presence of amplitude variability of the $\ell =1$ mode continues.
In the August 1996 data, we saw the most extreme example yet, where
all of the observed light was in a single ($k=8$ mode) for one night
(which we call the ``forte''). Data before and after the run show
power in the nights before and after in other pulsation modes besides
the $k=8$ and a much lower amplitude.
The periods from the 1996 data are consistent with the 2000 data set,
although there are differences in the details.

\section{Discussion and Puzzles}

Using the
$k=7$, 19, and 20 modes in seismological fits produces a best-fitting
model that is similar to that derived from only the $k=8$ through 18
modes, indicating that the new modes do not deviate drastically from
the expected mode pattern.

The reappearance of modes with frequencies similar to those obtained
before the mode disappeared (true of all modes from $k=8$ through 19),
shows that the stellar structure sampled by these modes remained the
same for almost 20 years. This is in spite of rapid amplitude change
events like the ``forte'' one observed in August 1996. Our
observations, coupled with guidance from the available theories of
Brickhill (1992) and Wu {\&} Goldreich (2001) suggest that the ``forte''
event was probably an extreme manifestation of a nonlinear mode-coupling event
that did not materially affect the structure of the star other than
possibly the driving region. The appearance and disappearance of modes is
similar to the behavior observed in the ZZ Ceti star G~29-38
by Kleinman et al. (1998), and we note that ``ensemble'' seismology
works for GD~358 as well as for the cool ZZ Ceti stars.
The one caveat is that the $\sim 1~\mu$Hz frequency ``wobbles'' will
place a limit on the accuracy of the seismology.

We also appear to have discovered an $\ell =2$ mode (at $1255.4~\mu$Hz)
in GD~358 for the first time, based on the match of the observed
period to that of $\ell =2$ modes from our best fitting model. Our
model indicates that this is the $k=34$ mode. This mode has a relatively
large amplitude of $14.9$~mma, which combined with the increased
geometric cancellation (about $3.8\times$) of an $\ell =2$ mode,
implies that it has the largest amplitude of any mode observed in 2000.
We note the existence of several linear combination peaks
involving the $1255~\mu$Hz mode, that also show complex structures.
This lends credence to the $1255~\mu$Hz mode being a real mode,
and that the complex structure is associated with the real mode
(such as amplitude modulation),
as opposed to being some sort of combination peak.
The amplitude of the $1255~\mu$Hz mode changed during the WET run,
so we suspect that the many subcomponents observed are most likely
due to amplitude modulation.

The period structure of the 1990 and 1994 WET data sets are similar, but
show that the amplitude of the modes, and even the fine structure,
changes with time. In August 1996, the period structure changed
rapidly and dramatically,
with essentially all the observed pulsation power going to the $k=8$
mode.
In spite of the large amplitude, the light curve was surprisingly
sinusoidal, with a small contribution from the $k=9$ mode.
Single site observations
one month earlier (June 1996) and one month later (September 1996)
show a period structure similar to those present in the 1990
and 1994 data sets. For the 2000 data set, the period structure shows
close to equal frequency splittings,
and the fine structure is different than observed before. Only the
$k=9$ mode show the same clear triplet observed in 1990 and 1994,
with the same frequency splitting.  The $k=8$ mode shows the $m=-1$ and
$m=1$ modes, while the central $m=0$ mode is below our
$4\langle {\rm Amp}\rangle$ significance level.
The other modes do not show clearly the triplet structure
previously observed.
The 1990 and 1994 data sets show the m-splitting
expected by rotational splitting, but the change of the splitting
frequency difference from 6~$\mu$Hz to 3~$\mu$Hz from $k=17$ to $k=8$
was interpreted as indicating differential rotation.

The apparent anticorrelation between the abundance of multiplet
structure and the highest order of combination frequencies seen is
a puzzle.
As we do not expect the differential rotation profile of GD~358
changed in the last 10 years (and the splittings we do see in 2000
support this contention), the anticorrelation must be telling something
about what is going on with rotation in the convection zone.
We say this because the combination peaks are believed to be caused
by the nonlinear response of the depth varying convection zone, and
thus the increased order of combination peaks implies that the convection
zone is more ``efficient'' at mixing eigenmodes to observable amplitudes.
The $k=8$ and 9 modes continue to show obvious multiplet structure and
little, if any, change in splitting. These modes are the most
``internal'' of the observed modes of GD~358, and we speculate that
this must have some bearing on their multiplet structure's ability to
persist.
We do not see any obvious pattern in the dominant amplitude multiplet
member with overtone number, so there is not an obvious pattern of
rotational coupling to the convection zone for determining mode
amplitude. We will need theoretical guidance to make sense of these
observations.

A related puzzle is the presence of extra multiplet members and/or
apparent large frequency shifts of modes in the $k=15$ and 16
multiplets.
The $k=15$ mode shows an extra component at $1430.88~\mu$Hz in the
1994 data and a peak at $1423.62~\mu$Hz in the 2000 data that have
not been seen before or since.
Some possible explanations include: rapid amplitude modulation of a
$k=15$ multiplet member that the FT interprets as an extra peak;
the 2000 peak is about the right frequency to be another $\ell =2$
mode, if we use the $1255.4~\mu$Hz mode as a reference point;
it could be an unattributed combination peak involving sums and
differences of known modes; or it could be something else entirely.
The large peak at about $1379~\mu$Hz in 1996 and 2000 is also something
of a mystery. It is possible that the $k=16$, $m=-1$ component
really changed by $10~\mu Hz$ from the $1368~\mu$Hz observed in 1990,
although we would have to explain why only this large amplitude multiplet
member suffered this large a frequency change.
Other possibilities include: the peak is a 1 cycle per day alias of
another mode; the peak is a combination peak --- the combination
$15+({\ell =2})-17$ is a perfect frequency match; or possibly an $\ell =2$
mode, based on period spacing arguments.
Further observations, data analysis with tools like wavelet analysis,
and further model fitting may help determine which explanation fits the
data best.

Brickhill (1992) proposed that the combination frequencies result from mixing
of the eigenmode signals by a depth-varying surface convection zone when
undergoing pulsation.
He pointed out that the convective turnover time in DA and DB variable
white dwarf stars occurs on a timescale much shorter than the pulsation
period. As a consequence, the convective region adjusts almost
instantaneously during a pulsation cycle. Brickhill demonstrated
that the flux leaving the convective zone depends on the depth
of the convective zone, which changes during the pulsation cycle,
distorting the observed flux. This distortion introduces combination
frequencies, even if the pulsation at the bottom of the convection zone
is linear, i.e., a single sinusoidal frequency.
Wu (2001) analytically calculated the amplitude and phases expected
of such combination frequencies, and concluded that the convective
induced distortion was roughly in agreement with GD358's 1994 observations,
provided that the inclination of the pulsation axis to the
line of sight is between $40\deg$ and $50\deg$.
Wu also calculated that the harmonics for $\ell=2$ modes should be much
higher than for $\ell=1$. However the theory overpredicts the amplitude
of the $\ell =1$ harmonics. She also predicts that
geometrical cancellation  will, in principle, allow a determination of
$\ell$ if both frequencies sums and differences are observed.
These predictions still need testing.

While Wu {\&} Goldreich (2001) discuss parametric instability mechanisms
for the amplitude of the pulsation modes, they only discuss the case
where the parent mode is unstable and the daughter modes are stable.
However, with GD~358, we have a different situation. The highest frequency
$k=8$ and 9 modes can have as a daughter mode one of the lower frequency
($k=17$, 18, or 19) $\ell =1$ modes and an $\ell =$~higher mode. One or
both or these daughter modes are actually pulsationally unstable as well,
which we believe would require coupling to still lower frequency
granddaughter modes that are predicted to be stable by our models and
the calculations of Brickhill (1990, 1991) and Goldreich {\&} Wu (1999a, b).
We suggest that occasionally the nonlinear coupling of the granddaughter
and daughter modes with the $k=8$ and 9 modes can allow the $k=8$ and
9 modes to suffer abrupt amplitude changes when everything is ``just
right''. In the meantime, the granddaughter modes will couple to the
excited daughter modes ($k=13$ through 19 in general) to produce the
observed amplitude instability  of these modes. We need a quantitative
theoretical treatment of this circumstance worked out to see if the
predicted behavior matches what we observe in GD~358.

Observations of GD~358 have been both rewarding and vexing. We have been
rewarded with enough $\ell =1$ modes being present to decipher the mode
structure and perform increasingly refined asteroseismology of this star,
starting with Bradley {\&} Winget (1994) up to the latest paper of Metcalfe
et~al. (2002). One thing asteroseismology has not provided us with is the
structure of and/or the depth of the surface convection zone. This would
help us test the ``convective driving'' mechanism introduced by Brickhill
(1991) and elaborated on by Goldreich {\&} Wu (1999a,b).
Our observations point out the need for further refinements of the
parametric instability mechanism described by Wu {\&} Goldreich (2001)
to better cover the observed mode behavior. The observational data set
is quite rich, and coupled with more detailed theories, offers the
promise of being able to unravel the mysteries of amplitude variation
seen in the DBV and DAV white dwarfs. This in turn, may offer us the
insights needed to ascertain why only some of the predicted modes are
seen at any one time.

MAW, AKJ, AEC, and MLB acknowledge
support by the National Science Foundation through the Research
Experiences for Undergraduates Summer Site Program to Florida Tech.

\begin{table}
\caption{Linear Combination of Peaks in 2000\label{combination}}
\begin{tabular}{c c c c c c c}
  k          &
  Period     &
  Amp.    &
  $f_{obs}$  &
  Combination&
  $f_{comb}$ &
  $\Delta f = f_{obs}-f_{comb}$\\
  &
  (s)&
  (mma)&
  ($\mu$Hz)&
  &
  ($\mu$Hz)&
  ($\mu$Hz)\\
$   $  &  $8096.02$  &  $3.518$  &  $ 123.52$  & $ 15- 17$  &  $ 123.49$  &  $
0.02$  \\
  ''      &    ''      &    ''      &    ''      &  $ 16-\ell=2$  & 
$ 123.38$  &
  $ 0.14$  \\
  ''      &    ''      &    ''      &    ''      &  $ 17- 19$  &  $ 123.61$  & 
$-0.09$  \\
$   $  &  $6078.39$  &  $2.169$  &  $ 164.52$  & $ 15- 18$  &  $ 164.68$  & 
$-0.17$  \\
$   $  &  $6032.23$  &  $1.340$  &  $ 165.78$  & $15-18^f$   & 165.65  & 0.12 
\\
$   $  &  $5669.64$  &  $1.152$  &  $ 176.38$  &    &   &   \\
$   $  &  $1765.91$  &  $1.738$  &  $ 566.28$  & $ 11- 17$  &  $ 566.27$  &  $
0.01$  \\
$   $  &  $1539.52$  &  $1.009$  &  $ 649.56$  &    &   &   \\
$   $  &  $1450.02$  &  $2.140$  &  $ 689.64$  & $ 11- 19$  &  $ 689.88$  & 
$-0.23$  \\
$   $  &  $1369.15$  &  $3.584$  &  $ 730.38$  & $ 10- 17$  &  $ 730.40$  & 
$-0.02$  \\
$   $  &  $1289.88$  &  $0.881$  &  $ 775.26$  & $9^0- 16$  &  $ 775.23$  &  $
0.03$  \\
$   $  &  $1166.29$  &  $2.876$  &  $ 857.42$  & $9^0- 17$  &  $ 857.42$  &  $
0.00$  \\
$   $  &  $1064.99$  &  $3.162$  &  $ 938.97$  & $8^+- 15$  &  $ 939.02$  & 
$-0.04$  \\
$   $  &  $1056.84$  &  $1.121$  &  $ 946.22$  & $8^-- 15$  &  $ 946.17$  &  $
0.04$  \\
$   $  &  $ 959.37$  &  $0.977$  &  $1042.35$  & $9^0- 20$  &  $1043.07$  & 
$-0.72$  \\
$   $  &  $ 941.28$  &  $0.895$  &  $1062.39$  & $8^+- 17$  &  $1062.51$  & 
$-0.12$  \\
$   $  &  $ 900.84$  &  $1.401$  &  $1110.07$  &  $8^--\ell=2^a$  & 
$1110.03$  &  $
0.05$  \\
$ 20$  &  $ 900.13$  &  $2.029$  &  $1110.95$  &  $8^-- \ell=2$ 
&  $1110.86$
  &  $ 0.10$  \\
$   $  &  $ 853.57$  &  $1.816$  &  $1171.55$  &    &   &   \\
$ 19$  &  $ 852.52$  &  $2.740$  &  $1172.99$  &    &   &   \\
$ \ell=2^h?$  &  $ 798.80$  &  $3.662$  &  $1251.87$  &$\ell=2-3.54\mu$Hz   
&   &   \\
$ \ell=2^g?$  &  $ 797.63$  &  $5.858$  &  $1253.72$  &    &   &   \\
$ \ell=2^f?$  &  $ 797.17$  &  $5.330$  &  $1254.44$  &    &   &   \\
$ \ell=2$  &  $ 796.55$  &  $14.870$  &  $1255.41$  &not 1235$\mu$Hz 
&   &  
\\
$ \ell=2^a?$  &  $ 796.02$  &  $7.508$  &  $1256.24$  &   &   &   \\
$ \ell=2^b?$  &  $ 795.73$  &  $1.280$  &  $1256.71$  &   &   &   \\
$ \ell=2^c?$  &  $ 795.36$  &  $3.277$  &  $1257.29$  & $\ell=2+1.88\mu$Hz   &
   &   \\
$\ell=2^d?$  &  $ 794.75$  &  $2.433$  &  $1258.26$  &18 
$\ell=2+2.85\mu$Hz    &
   &   \\
$\ell=2^e?$  &  $ 793.88$  &  $1.568$  &  $1259.63$  &18 
$\ell=2+4.22\mu$Hz    &
   &   \\
$   $  &  $ 782.89$  &  $1.546$  &  $1277.31$  &     &   &   \\
$   $  &  $ 781.92$  &  $1.350$  &  $1278.90$  &     &   &   \\
$17^a$  &  $ 771.68$  &  $1.221$  &  $1295.87$  &    &   &   \\
$ 17$  &  $ 771.25$  &  $27.940$  &  $1296.60$  &     &   &   \\
$17^b$  &  $ 770.80$  &  $1.604$  &  $1297.36$  &    &   &   \\
$    $  &  $759.39$   &  $1.205$  &  $1316.85$  &    &   &   \\
$   $  &  $ 725.70$  &  $1.286$  &  $1377.98$  & $ 7-17^b$  &  $1378.13$  & 
$-0.15$  \\
$ 16$  &  $ 725.27$  &  $5.157$  &  $1378.80$  & $15+18\ell=2-17$ & 1378.76  &
0.03  \\
$   $  &  $ 724.78$  &  $2.688$  &  $1379.73$  &    &   &   \\
$   $  &  $ 709.03$  &  $1.185$  &  $1410.38$  &    &   &   \\
$15^+$  &  $ 704.18$  &  $29.720$  &  $1420.10$  &    &   &   \\
$15^a$  &  $ 702.44$  &  $3.003$  &  $1423.62$  & 15+3.52$\mu$Hz   &   &   \\
$   $  &  $ 690.99$  &  $1.123$  &  $1447.21$  &    &   &   \\
$ 12$  &  $ 575.94$  &  $1.030$  &  $1736.29$  &    &   &   \\
$ 11$  &  $ 536.81$  &  $0.830$  &  $1862.87$  &    &   &   \\
$ 10$  &  $ 493.34$  &  $1.280$  &  $2027.00$  &    &   &   \\
$9^+$  &  $ 465.01$  &  $2.980$  &  $2150.49$  &9+3.54$\mu$Hz    &   &   \\
$9^0$  &  $ 464.25$  &  $5.300$  &  $2154.03$  &    &   &   \\
$9^-$  &  $ 463.45$  &  $2.510$  &  $2157.72$  &9-3.69$\mu$Hz    &   &   \\
$   $  &  $ 447.30$  &  $0.968$  &  $2235.66$  &$17+8^+-15^+$&2235.61&0.08\\
$   $  &  $ 439.08$  &  $0.989$  &  $2277.50$  &$15^++9-17$&2277.53&-0.02\\
\end{tabular}
\end{table}
\begin{table}
\begin{tabular}{c c c c c c c}
  k          &
  Period     &
  Amp.    &
  $f_{obs}$  &
  Combination&
  $f_{comb}$ &
  $\Delta f = f_{obs}-f_{comb}$\\
  &
  (s)&
  (mma)&
  ($\mu$Hz)&
  &
  ($\mu$Hz)&
  ($\mu$Hz)\\
$8^+$  &  $ 423.89$  &  $5.640$  &  $2359.11$  &    &   &   \\
$8^-$  &  $ 422.61$  &  $5.620$  &  $2366.27$  &$8^+-2\times 
3.58\mu$Hz &   &  
\\
$   $  &  $ 415.34$  &  $1.030$  &  $2407.65$  & $ 20+ 17$  &  $2407.56$  &  $
0.09$  \\
$   $  &  $ 405.15$  &  $1.388$  &  $2468.21$  & $ 19+ 17^a$  &  $2468.86$  & 
$-0.65$  \\
$   $  &  $ 404.89$  &  $1.091$  &  $2469.83$  &  $ 19+ 17$  & 
$2469.60$  &  $
0.24$  \\
$   $  &  $ 398.57$  &  $1.249$  &  $2508.96$  &  $  2 \times \ell=2^f$  & 
$2508.89$  &  $ 0.08$  \\
$   $  &  $ 398.29$  &  $1.423$  &  $2510.74$  &  $  2 \times  \ell=2$  & 
$2510.83$  &  $-0.09$  \\
$   $  &  $ 398.15$  &  $2.052$  &  $2511.65$  &  $  2 \times  \ell=2^a$  & 
$2512.49$  &  $-0.84$  \\
$   $  &  $ 392.10$  &  $1.670$  &  $2550.34$  & $\ell=2^g+ 17$  & 
$2550.32$  &
  $ 0.01$  \\
$   $  &  $ 392.00$  &  $1.737$  &  $2551.04$  &  $ \ell=2^g+ 17^b$  & 
$2551.08$  &  $-0.04$  \\
  ''      &    ''      &    ''      &    ''      &  $ \ell=2^f+ 17$ 
&  $2551.05$
  &  $ 0.00$  \\
  ''      &    ''      &    ''      &    ''      &  $ \ell=2+ 17^a$ 
&  $2551.28$
  &  $-0.24$  \\
$   $  &  $ 391.85$  &  $3.703$  &  $2552.02$  &  $ 19+ 16$  & 
$2551.79$  &  $
0.23$  \\
  ''      &    ''      &    ''      &    ''      &  $ \ell=2^f+ 17^a$  & 
$2551.80$  &  $ 0.22$  \\
  ''      &    ''      &    ''      &    ''      &  $ \ell=2+ 17$  & 
$2552.02$ 
&  $ 0.00$  \\
  ''      &    ''      &    ''      &    ''      &  $ \ell=2^a+ 17^a$  & 
$2552.11$  &  $-0.09$  \\
$   $  &  $ 391.72$  &  $2.007$  &  $2552.83$  & $ \ell=2+ 17^b$  & 
$2552.77$ 
&  $ 0.05$  \\
  ''      &    ''      &    ''      &    ''      &  $ \ell=2^a+ 17$ 
&  $2552.85$
  &  $-0.02$  \\
  ''      &    ''      &    ''      &    ''      &  $ \ell=2^b+ 17^a$  & 
$2552.58$  &  $ 0.24$  \\
  ''      &    ''      &    ''      &    ''      &  $ \ell=2^c+ 17$ 
&  $2553.32$
  &  $-0.49$  \\
  ''      &    ''      &    ''      &    ''      &  $ \ell=2^c+ 17^a$  & 
$2553.16$  &  $-0.34$  \\
$   $  &  $ 391.56$  &  $1.159$  &  $2553.88$  & $ \ell=2^a+ 17^b$ 
&  $2553.60$
  &  $ 0.28$  \\
  ''      &    ''      &    ''      &    ''      &  $ \ell=2^b+ 17^b$  & 
$2554.07$  &  $-0.19$  \\
  ''      &    ''      &    ''      &    ''      &  $ \ell=2^c+ 17$ 
&  $2553.90$
  &  $-0.02$  \\
  ''      &    ''      &    ''      &    ''      &  $ \ell=2^c+ 17^b$  & 
$2554.65$  &  $-0.77$  \\
  ''      &    ''      &    ''      &    ''      &  $ \ell=2^e+ 17^a$  & 
$2554.13$  &  $-0.25$  \\
$   $  &  $ 385.62$  &  $7.759$  &  $2593.21$  &  $  2 \times  17$ 
&  $2593.21$
  &  $ 0.00$  \\
$   $  &  $ 379.48$  &  $0.928$  &  $2635.16$  &  $ \ell=2+ 16$  & 
$2634.21$  &
  $ 0.95$  \\
  ''      &    ''      &    ''      &    ''      &  $ \ell=2^a+ 16$ 
&  $2635.04$
  &  $ 0.12$  \\
  ''      &    ''      &    ''      &    ''      &  $ \ell=2^b+ 16$ 
&  $2635.51$
  &  $-0.35$  \\
$   $  &  $ 374.24$  &  $1.576$  &  $2672.05$  &  $19^a+ 15$  & 
$2671.97$  &  $
0.08$  \\
$   $  &  $ 374.00$  &  $3.459$  &  $2673.81$  &  $\ell=2^g+ 15$  & 
$2673.82$ 
&  $-0.01$  \\
$   $  &  $ 373.89$  &  $3.498$  &  $2674.56$  &  $\ell=2^f+ 15$  & 
$2674.54$ 
&  $ 0.02$  \\
  ''      &    ''      &    ''      &    ''      &  $ 17^a+ 16$  & 
$2674.67$  &
$-0.10$  \\
$ 7$  &  $ 373.76$  &  $8.430$  &  $2675.49$  &  $ 17+ 16$  &  $2675.40$  &  $
0.09$  \\
$   $  &  $ 373.64$  &  $4.254$  &  $2676.40$  &  $\ell=2^a+ 15$  & 
$2676.34$ 
&  $ 0.06$  \\
  ''      &    ''      &    ''      &    ''      &  $\ell=2^b+ 15$  & 
$2676.81$ 
&  $-0.41$  \\
  ''      &    ''      &    ''      &    ''      &  $ 17^b+ 16$  & 
$2676.15$  & 
$ 0.25$  \\
$   $  &  $ 373.50$  &  $1.827$  &  $2677.38$  &  $\ell=2^g+15^a$  & 
$2677.34$ 
&  $ 0.04$  \\
  ''      &    ''      &    ''      &    ''      &  $ \ell=2^c+ 15$ 
&  $2677.39$
  &  $-0.01$  \\
$   $  &  $ 373.38$  &  $1.193$  &  $2678.22$  &  $ \ell=2^g+ 15^a$  & 
$2678.06$  &  $ 0.16$  \\
  ''      &    ''      &    ''      &    ''      &  $ \ell=2^e+ 15$ 
&  $2678.35$
  &  $-0.13$  \\
$   $  &  $ 373.16$  &  $0.870$  &  $2679.78$  &  $ \ell=2+ 15^a$  & 
$2679.03$ 
&  $ 0.75$  \\
  ''      &    ''      &    ''      &    ''      &  $ \ell=2^a+ 15^a$  & 
$2679.86$  &  $-0.08$  \\
  ''      &    ''      &    ''      &    ''      &  $ \ell=2^b+ 15^a$  & 
$2680.33$  &  $-0.55$  \\
$   $  &  $ 372.83$  &  $0.932$  &  $2682.18$  & $15^a+\ell=2^d$  & 2681.87  &
0.30  \\
$   $  &  $ 368.09$  &  $5.653$  &  $2716.71$  &  $ 17+ 15$  & 
$2716.70$  &  $
0.01$  \\
$   $  &  $ 367.62$  &  $0.840$  &  $2720.17$  &  $ 17^a+ 15^a$  & 
$2719.49$  &
  $ 0.68$  \\
  ''      &    ''      &    ''      &    ''      &  $ 17+ 15^a$  & 
$2720.22$  & 
$-0.05$  \\
$   $  &  $ 357.29$  &  $1.731$  &  $2798.86$  &  $ 16+ 15$  &  $2798.89$  & 
$-0.03$  \\
$   $  &  $ 357.16$  &  $0.877$  &  $2799.88$  &  $ 16^a+ 15$  & 
$2799.83$  & 
$ 0.05$  \\
\end{tabular}
\end{table}
\begin{table}
\begin{tabular}{c c c c c c c}
  k          &
  Period     &
  Amp.    &
  $f_{obs}$  &
  Combination&
  $f_{comb}$ &
  $\Delta f = f_{obs}-f_{comb}$\\
  &
  (s)& 
  (mma)&
  ($\mu$Hz)&
  &
  ($\mu$Hz)&
  ($\mu$Hz)\\ 
$   $  &  $ 352.09$  &  $4.260$  &  $2840.20$  &  $  2 \times  15$ 
&  $2840.19$
  &  $ 0.01$  \\
$   $  &  $ 351.66$  &  $1.054$  &  $2843.68$  &  $ 15+ 15a$  &  $2843.71$  & 
$-0.03$  \\
$   $  &  $ 293.60$  &  $0.992$  &  $3405.99$  &  $ \ell=2+9^+$  & 
$3405.91$  &
  $ 0.09$  \\
  ''      &    ''      &    ''      &    ''      &  $\ell=2^a+9^+$  & 
$3406.74$ 
&  $-0.74$  \\
  ''      &    ''      &    ''      &    ''      &  $ 16+ 10$  & 
$3405.79$  &  $
0.20$  \\
$   $  &  $ 289.80$  &  $1.929$  &  $3450.64$  &  $ 17^a+ 9^0$  & 
$3449.90$  & 
$ 0.74$  \\
  ''      &    ''      &    ''      &    ''      &  $ 17+ 9^0$  & 
$3450.63$  & 
$ 0.01$  \\
  ''      &    ''      &    ''      &    ''      &  $ 15^a+ 10$  & 
$3450.62$  & 
$ 0.02$  \\
$   $  &  $ 279.78$  &  $1.393$  &  $3574.19$  &  $ 15+9^0$  & 
$3574.12$  &  $
0.07$  \\
  ''      &    ''      &    ''      &    ''      &  $15^a+9^+$  & 
$3574.11$  & 
$ 0.08$  \\
$   $  &  $ 279.50$  &  $1.118$  &  $3577.77$  & $ 15+9^-$  &  $3577.82$  & 
$-0.05$  \\
  ''      &    ''      &    ''      &    ''      &  $15^a+9^0$  & 
$3577.65$  & 
$ 0.13$  \\
$   $  &  $ 276.66$  &  $1.550$  &  $3614.54$  & $\ell=2 + 8^+$  & 
$3614.53$  &
  $ 0.01$  \\
$   $  &  $ 276.10$  &  $0.960$  &  $3621.85$  & $ \ell=2+ 8^-$  & 
$3621.68$  &
  $ 0.17$  \\
$   $  &  $ 273.54$  &  $1.284$  &  $3655.77$  & $ 17+ 8^+$  & 
$3655.72$  &  $
0.05$  \\
$   $  &  $ 273.01$  &  $1.382$  &  $3662.91$  &  $ 17+8^-$  & 
$3662.87$  &  $
0.03$  \\
$   $  &  $ 264.60$  &  $1.523$  &  $3779.25$  &  $ 15+8^+$  & 
$3779.21$  &  $
0.04$  \\
$   $  &  $ 264.11$  &  $3.649$  &  $3786.37$  &  $ 15+8^-$  & 
$3786.37$  &  $
0.00$  \\
  ''      &    ''      &    ''      &    ''      &  $ 20+7$  &  $3786.44$  & 
$-0.07$  \\
$   $  &  $ 259.84$  &  $1.569$  &  $3848.54$ &  $ 19+ 7$  &  $3848.48$  &  $
0.06$  \\
$   $  &  $ 259.78$  &  $0.885$  &  $3849.36$  &$2\times
17+\ell=2$&3848.61&0.75\\
$   $  &  $ 257.08$  &  $3.204$  &  $3889.80$  &  $  3 \times  17$ 
&  $3889.81$
  &  $-0.02$  \\
$   $  &  $ 254.51$  &  $1.744$  &  $3929.10$  & $ \ell=2^g+ 7$  & 
$3929.21$  &
  $-0.11$  \\
  ''      &    ''      &    ''      &    ''      &  $\ell=2+ \ell=2^g+15$  & 
$3929.22$  &  $-0.12$  \\
  ''      &    ''      &    ''      &    ''      &  $ 15+ 2\times \ell=2^f$  & 
$3929.06$  &  $ 0.04$  \\
$   $  &  $ 254.42$  &  $1.140$  &  $3930.46$  & $ \ell=2^f+ 7$  & 
$3929.93$  &
  $ 0.53$  \\
  ''      &    ''      &    ''      &    ''      &  $ 16+ \ell=2+17^a$  & 
$3929.84$  &  $ 0.62$  \\
  ''      &    ''      &    ''      &    ''      &  $ 15+ 2\times \ell=2$  & 
$3930.83$  &  $-0.37$  \\
$   $  &  $ 254.34$  &  $2.986$  &  $3931.77$  &  $ 2\times 17+\ell=2+16$  & 
$3931.76$  &  $ 0.00$  \\
$   $  &  $ 251.82$  &  $1.015$  &  $3971.17$  &  $ 17+ \ell=2^g+15$  & 
$3970.41$  &  $ 0.75$  \\
  ''      &    ''      &    ''      &    ''      &  $ 15+ \ell=2^f+17$  & 
$3970.44$  &  $ 0.73$  \\
  ''      &    ''      &    ''      &    ''      &  $ 15+ \ell=2+17^a$  & 
$3971.14$  &  $ 0.03$  \\
$   $  &  $ 251.76$  &  $2.242$  &  $3972.03$  &  $ 15+ 17+\ell=2$ 
&  $3972.12$
  &  $-0.08$  \\
  ''      &    ''      &    ''      &    ''      &  $ 16+ 2\times 17$  & 
$3972.01$  &  $ 0.03$  \\
  ''      &    ''      &    ''      &    ''      &  $ 17+ 7$  &  $3972.09$  & 
$-0.06$  \\
$   $  &  $ 251.70$  &  $1.395$  &  $3972.93$  &  $ 17+ \ell=2^a+15$  & 
$3973.01$  &  $-0.07$  \\
  ''      &    ''      &    ''      &    ''      &  $ 15+ 17^a+16$  & 
$3972.92$ 
&  $ 0.01$  \\
$   $  &  $ 249.17$  &  $1.822$  &  $4013.33$  &  $ 2\times 17+ 
2\times 15$  & 
$4013.31$  &  $ 0.02$  \\
$   $  &  $ 246.59$  &  $0.948$  &  $4055.26$  &  $ 15+ 16+\ell=2$ 
&  $4055.26$
  &  $ 0.00$  \\
$   $  &  $ 244.27$  &  $1.462$  &  $4093.91$  &  $ 15+ 2\times \ell=2$  & 
$4093.90$  &  $ 0.01$  \\
$   $  &  $ 244.22$  &  $1.576$  &  $4094.67$  &  $ \ell=2^g+ 2\times 15$  & 
$4094.64$  &  $ 0.03$  \\
  ''      &    ''      &    ''      &    ''      &  $ 17^a+ 16+15$  & 
$4094.73$ 
&  $-0.06$  \\
  ''      &    ''      &    ''      &    ''      &  $ 2\times 15+ \ell=2^f$  & 
$4094.66$  &  $ 0.02$  \\
$   $  &  $ 244.16$  &  $3.487$  &  $4095.59$  &  $ \ell=2+ 2\times 15$  & 
$4095.61$  &  $-0.02$  \\
  ''      &    ''      &    ''      &    ''      &  $ 17+ 16+15$  & 
$4095.47$  &
  $ 0.12$  \\
  ''      &    ''      &    ''      &    ''      &  $ 15+ 7$  & 
$4095.58$  &  $
0.01$  \\
$   $  &  $ 244.11$  &  $1.722$  &  $4096.45$  &  $ 17+ 15+16^a$  & 
$4096.48$ 
&  $-0.03$  \\
  ''      &    ''      &    ''      &    ''      &  $ 16+ 17^a+16$  & 
$4096.50$ 
&  $-0.05$  \\
$   $  &  $ 210.65$  &  $1.196$  &  $4747.23$  &  $ 2\times 17+ 9^0$  & 
$4747.24$  &  $-0.01$  \\
$   $  &  $ 198.63$  &  $1.272$  &  $5034.61$  &  $ 15+ \ell=2+8^+$  & 
$5034.64$  &  $-0.03$  \\
$   $  &  $ 198.34$  &  $1.451$  &  $5041.82$  &  $ 15+ \ell=2+8^-$  & 
$5041.95$  &  $-0.14$  \\
\end{tabular}
\end{table}
\begin{table}
\begin{tabular}{c c c c c c c}
  k          &
  Period     &
  Amp.    &
  $f_{obs}$  &
  Combination&
  $f_{comb}$ &
  $\Delta f = f_{obs}-f_{comb}$\\
  &
  (s)& 
  (mma)&
  ($\mu$Hz)&
  &
  ($\mu$Hz)&
  ($\mu$Hz)\\ 
$   $  &  $ 196.74$  &  $0.992$  &  $5082.98$  &  $ 17+ 15+8^-$  & 
$5082.97$  &
  $ 0.00$  \\
$   $  &  $ 192.81$  &  $1.319$  &  $5186.46$  &  $  4 \times  17$ 
&  $5186.42$
  &  $ 0.04$  \\
  ''      &    ''      &    ''      &    ''      &  $ 3\times \ell=2+ 15$  & 
$5185.88$  &  $ 0.59$  \\
  ''      &    ''      &    ''      &    ''      &  $ 2\times \ell=2+ 
16+17$  & 
$5187.18$  &  $-0.72$  \\
$   $  &  $ 192.07$  &  $1.641$  &  $5206.47$  &  $ 2\times 15+ 8^-$  & 
$5206.47$  &  $ 0.00$  \\
$   $  &  $ 189.81$  &  $1.171$  &  $5268.55$  &  $ 2\times 17+ 7$ 
&  $5267.77$
  &  $ 0.78$  \\
  ''      &    ''      &    ''      &    ''      &  $ 2\times 17+ 
15+\ell=2$  & 
$5268.64$  &  $-0.09$  \\
$   $  &  $ 186.94$  &  $1.153$  &  $5349.20$  &  $ 3\times \ell=2+15$  & 
$5349.33$  &  $-0.12$  \\
$   $  &  $ 186.85$  &  $2.004$  &  $5351.84$  &  $  2 \times  7$  & 
$5350.97$ 
&  $ 0.87$  \\
  ''      &    ''      &    ''      &    ''      &  $ 15+16+17+\ell=2$  & 
$5351.86$  &  $-0.02$  \\
$   $  &  $ 185.46$  &  $0.842$  &  $5392.02$  &  $ 2\times 15+ 17+\ell=2$  & 
$5392.13$  &  $-0.11$  \\
$   $  &  $ 181.30$  &  $0.609$  &  $5515.65$  &  $ 3\times 15+\ell=2$  &
$5515.69$  &  $-0.04$  \\
$   $  &  $ 165.46$  &  $0.533$  &  $6043.85$  &  $ 15+17+\ell=2+9^0$  & 
$6043.83$  &  $ 0.01$  \\
$   $  &  $ 162.72$  &  $0.584$  &  $6145.42$  &  $ 8^++8^-+ 15$  & 
$6145.48$ 
&  $-0.07$  \\
$   $  &  $ 158.96$  &  $0.554$  &  $6290.83$  &  $ 2\times \ell=2+15+8^+$  & 
$6290.02$  &  $ 0.81$  \\
  ''      &    ''      &    ''      &    ''      &  $ 8^++ 16+17+\ell=2$  & 
$6290.88$  &  $-0.05$  \\
$   $  &  $ 154.76$  &  $0.958$  &  $6461.78$  &  $  3 \times  9^0$  & 
$6462.08$  &  $-0.30$  \\
  ''      &    ''      &    ''      &    ''      &  $ \ell=2+2\times 
15+8^-$  & 
$6461.88$  &  $-0.10$  \\
  ''      &    ''      &    ''      &    ''      &  $ 16+15+17+8^-$ 
&  $6461.77$
  &  $ 0.01$  \\
$   $  &  $ 147.67$  &  $0.761$  &  $6771.90$  &  $ 2\times 
\ell=2+3\times 15$ 
&  $6771.06$  &  $ 0.84$  \\
  ''      &    ''      &    ''      &    ''      &  $ 2\times 
\ell=2+15+16+17$  &
  $6771.94$  &  $-0.04$  \\
$   $  &  $ 129.69$  &  $0.438$  &  $7710.99$  &  $ 2\times 15+8^++8^-$  & 
$7710.93$  &  $ 0.06$  \\
  ''      &    ''      &    ''      &    ''      &  $ 8^++15+16+17+\ell=2$  & 
$7710.95$  &  $ 0.04$  \\
  ''      &    ''      &    ''      &    ''      &  $ 7+8^++15+\ell=2$  & 
$7710.10$  &  $ 0.89$  \\
$   $  &  $ 129.61$  &  $0.410$  &  $7715.61$  &  $ 8^-+3\times \ell=2+15$  & 
$7715.47$  &  $ 0.14$  \\
$   $  &  $ 126.87$  &  $0.502$  &  $7881.92$  &  $ 3\times 15+\ell=2+8^+$  & 
$7881.87$  &  $ 0.05$  \\
$   $  &  $ 125.87$  &  $0.410$  &  $7944.98$  &  $  2 \times 
15+2\times \ell=2+
2\times 17$  &  $7944.07$  &  $ 0.91$  \\
$   $  &  $ 124.56$  &  $0.401$  &  $8028.10$  &  $ 3\times 
\ell=2+3\times 15$ 
&  $8027.31$  &  $ 0.78$  \\
  ''      &    ''      &    ''      &    ''       & $ 3\times 
\ell=2+15+16+17$  &
  $8027.31$  &  $ 0.78$  \\
\end{tabular}
\end{table}

\clearpage
\clearpage
\clearpage
\clearpage
\clearpage
\clearpage

\clearpage
\clearpage
\begin{figure}
\centering
\includegraphics[width=\linewidth]{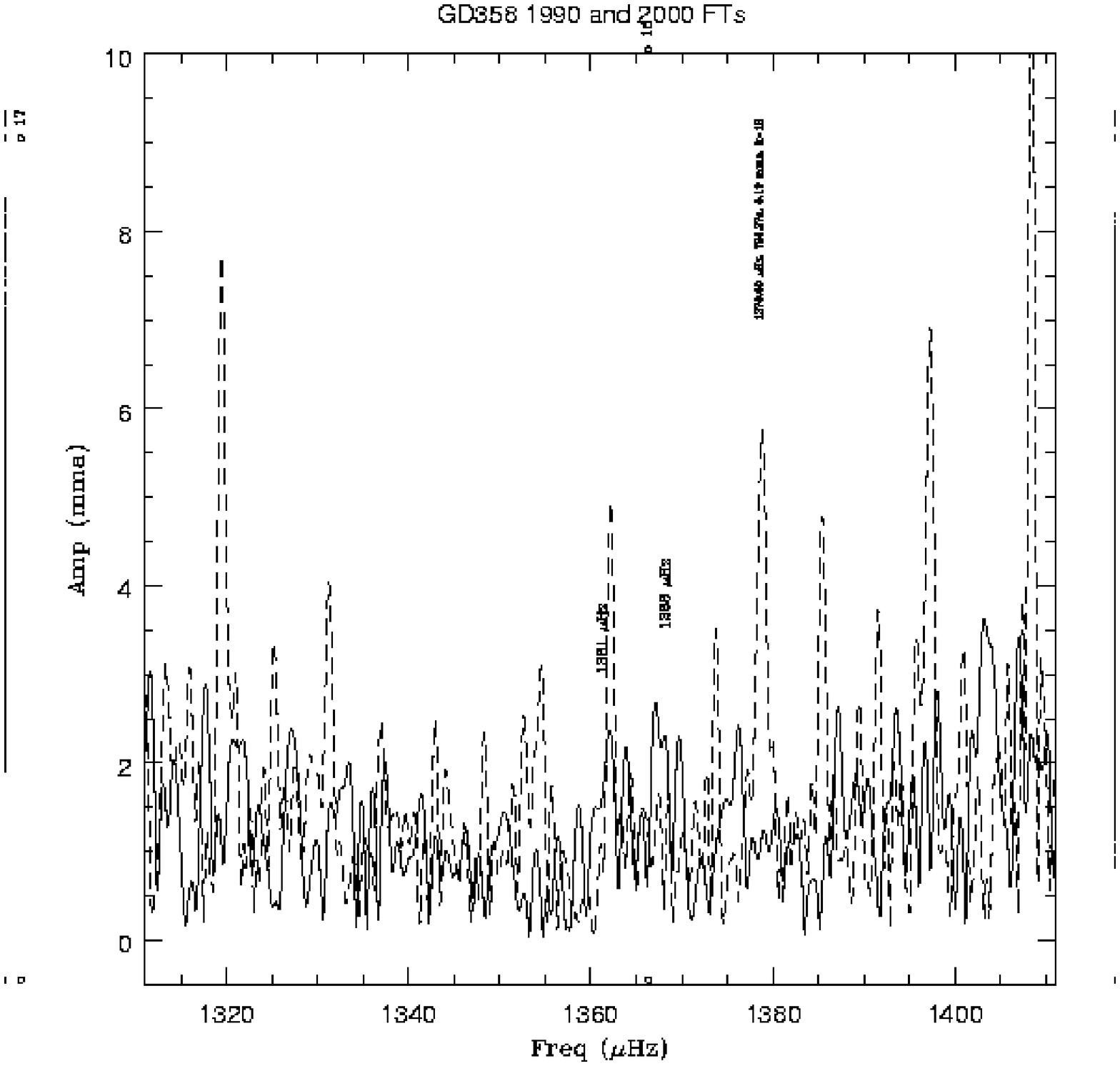}
\caption{Peaks around k=16 in the 1990 (solid line) and 2000 (dashed line)
transforms}
\label{k16}
\end{figure}

\clearpage
\begin{figure}
\centering
\includegraphics[width=\linewidth]{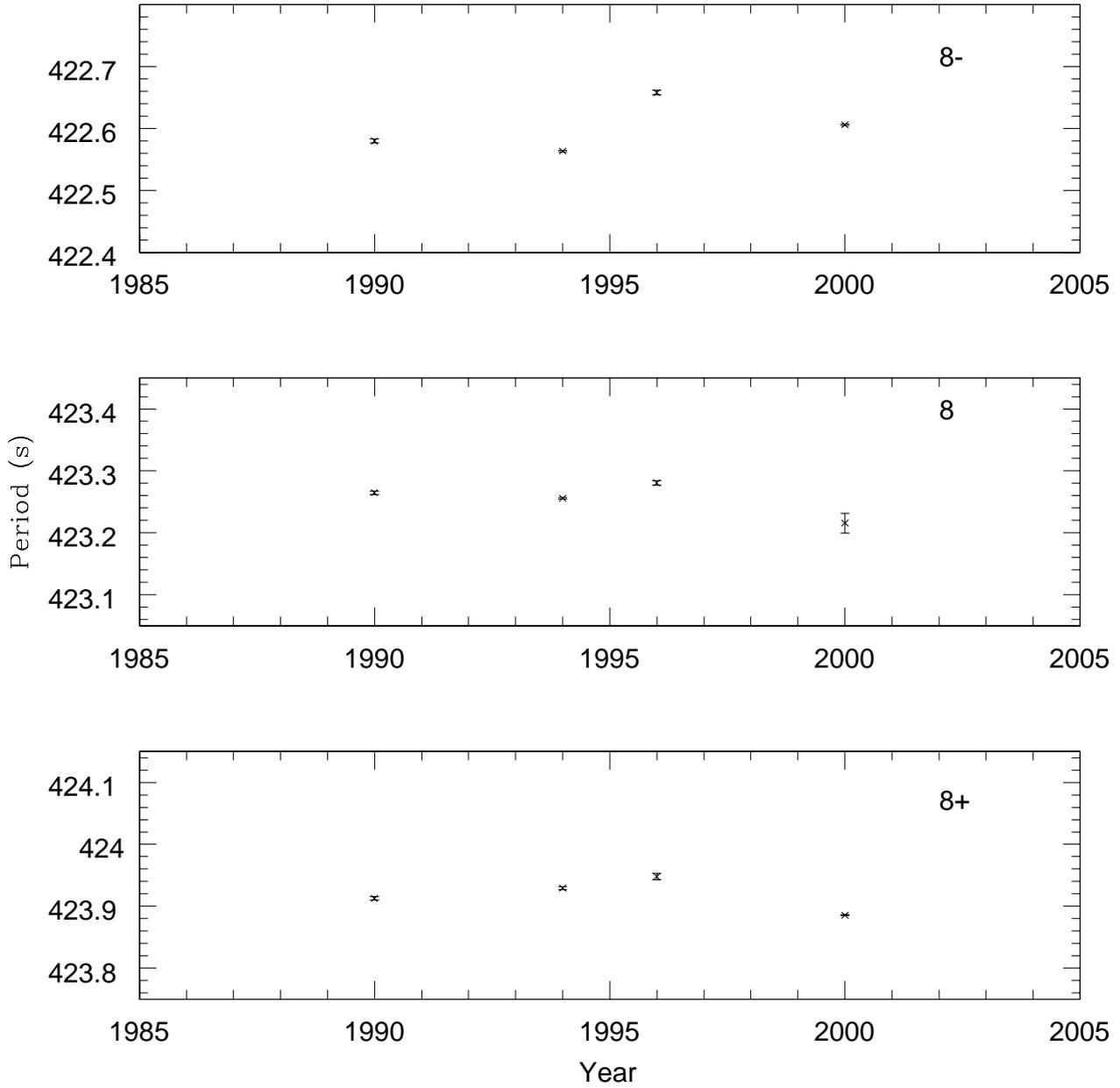}
\caption{Search for $\dot P$: The periods of the $m$
subcomponents of even the most stable mode,
$k=8$, change significantly, from year to year. The same
behavior is detected for all pulsations.}
\label{deltap8}
\end{figure}
\clearpage
\begin{figure}
\centering
\includegraphics[width=\linewidth]{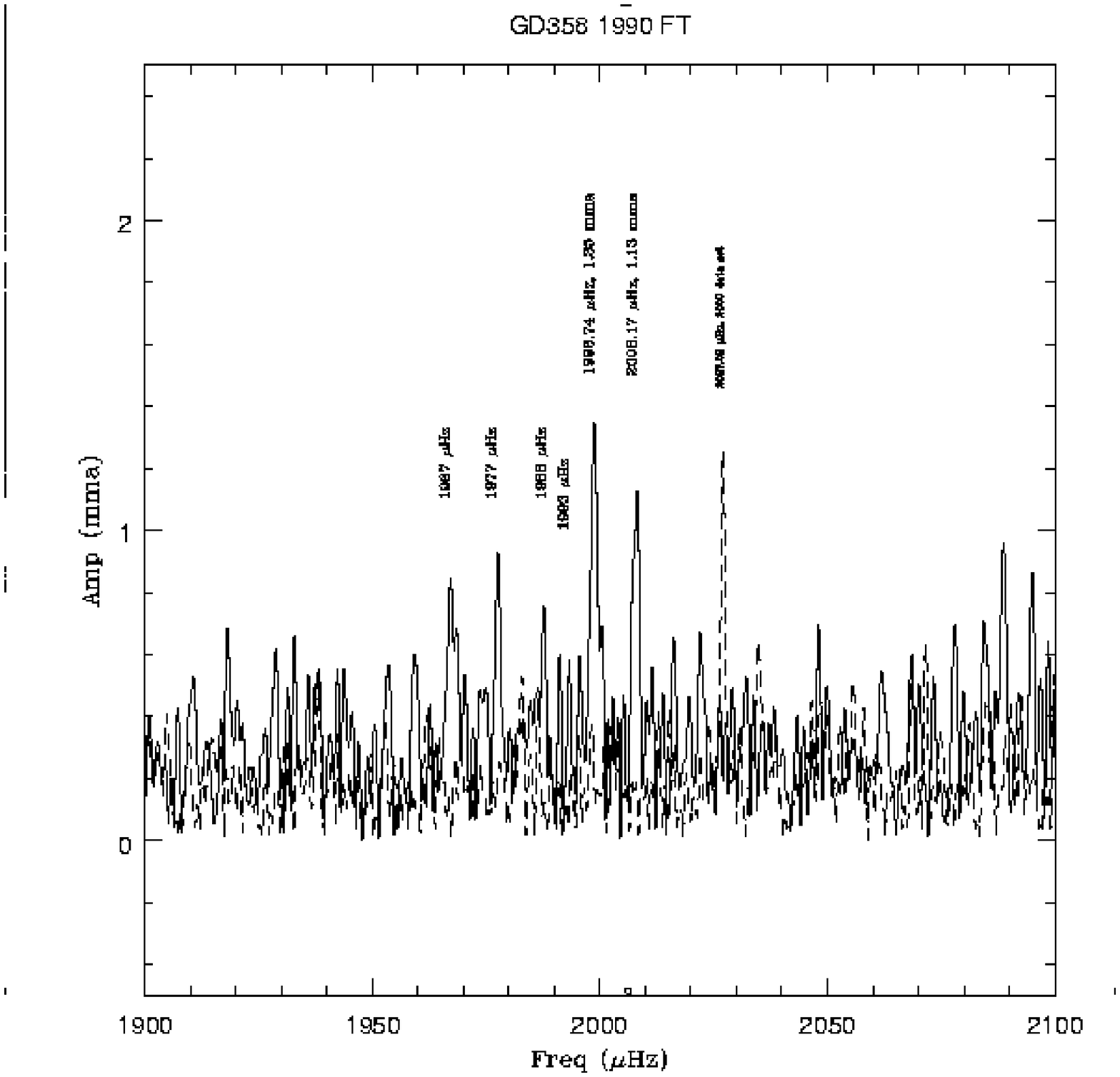}
\caption{Peaks around k=10 in the 1990 (solid line) and 2000 (dashed line)
transforms}
\label{k10}
\end{figure}
\clearpage
\begin{figure}
\centering
\includegraphics[width=\linewidth]{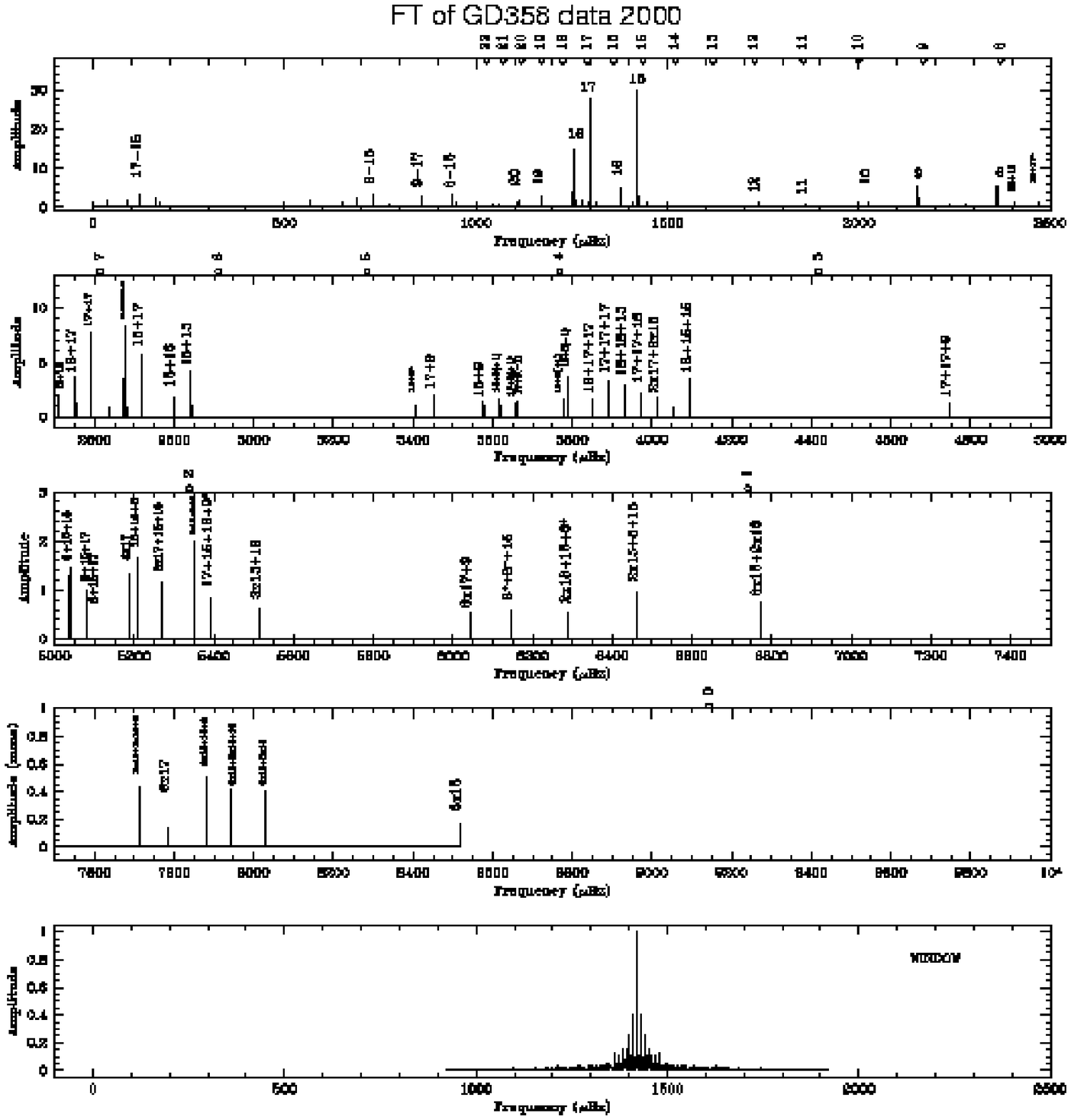}
\caption{Pre-whitened peaks in the 2000 transform}
\label{pk2000}
\end{figure}
\end{document}